\documentclass[12pt, draftclsnofoot, onecolumn]{IEEEtran}

\usepackage{amsmath}
\usepackage{amsfonts}
\usepackage{mathrsfs}
\usepackage{amssymb}
\usepackage{mathtools}
\usepackage{bm}
\usepackage{mathrsfs}
\usepackage{cite}

\usepackage[caption=false,font=footnotesize]{subfig}
\usepackage{graphicx}
\usepackage{stfloats}
\usepackage{acronym}
\usepackage{hyperref}

\newtheorem{lemma}{Lemma}
\newtheorem{corollary}{Corollary}
\newtheorem{theorem}{Theorem}

\allowdisplaybreaks
\newcommand{\bs}{ {\scriptscriptstyle\mathsf{BS}} }
\newcommand{\ue}{ {\scriptscriptstyle\mathsf{UE}} }
\newcommand{\intra}{\mathsf{intra}}
\newcommand{\inter}{\mathsf{inter}}
\newcommand{\LT}[2]{\mathcal{L}_{#1}\left(#2\right)}

\newcommand{\sinr}{\text{SINR}}

\acrodef{d2d}[D2D]{device-to-device}
\acrodef{mmwave}[mmWave]{millimeter wave}
\acrodef{bs}[BS]{base station}
\acrodef{ue}[UE]{user equipments}

\begin{document}
\title{Coverage Analysis of Integrated Sub-6GHz-mmWave Cellular Networks \\with Hotspots}

\author{Minwei Shi, Kai Yang,~\IEEEmembership{Member,~IEEE}, Zhu Han,~\IEEEmembership{Fellow,~IEEE}, \\and Dusit Niyato,~\IEEEmembership{Fellow,~IEEE}
\thanks{M. Shi and K. Yang are with the School of Information and Electronics, Beijing Institute of Technology, Beijing, China, and also with Beijing Key Laboratory of Fractional Signals and Systems, Beijing, China (email: yangkai@ieee.org).

Zhu Han is with the Department of Electrical and Computer Engineering, University of Houston, Houston, USA (e-mail: zhan2@uh.edu), and also with the Department of Computer Science and Engineering, Kyung Hee University, Seoul, South Korea.

Dusit Niyato is with the School of Computer Science and Engineering, Nanyang Technological University, Singapore (email: dniyato@ntu.edu.sg).}
}

\maketitle

\vspace{-0.8cm}

\begin{abstract}
Deploying Sub-6GHz networks together with millimeter wave (mmWave) is a promising solution to achieve high data rates in traffic hotspots while guaranteeing sufficient coverage, where mmWave small cells are densely deployed to provide high quality of service. In this paper, we propose an analytical framework to investigate the integrated Sub-6GHz-mmWave cellular networks, in which the Sub-6GHz base stations (BSs) are modeled as a Poisson point process, and the mmWave BSs are clustered following a Poisson cluster process in traffic hotspots. We conduct stochastic geometry-based analysis and derive the performance metrics including the association probability, signal-to-interference-plus-noise ratio coverage probability and average achievable rate, which are validated to be accurate by Monte Carlo simulations. We analyze the impact of various deployment parameters on the network performance to give insights on the network design. In particular, it is shown that deploying mmWave small cells in traffic hotspots will outperform both traditional Sub-6GHz heterogeneous network and isolated mmWave system in terms of the coverage probability. It can also be shown that extremely high and extremely small association weight for mmWave BSs will deteriorate the performance for cell edge users and cell interior users, respectively. Moreover, there exists an optimal pre-decided dispersion parameter of mmWave BSs that contributes to the maximum coverage probability.
\end{abstract}

\begin{IEEEkeywords}
Heterogeneous cellular networks, Sub-6GHz, millimeter wave, Poisson point process, Poisson cluster process.
\end{IEEEkeywords}

\section{Introduction}\label{sec:Introduction}
Millimeter wave~(mmWave) has been considered as a key technology to meet the ever-growing demand for mobile data rate due to its large available bandwidth \cite{Access2013Rappaport}. Yet it is challengeable to achieve the universal coverage with only mmWave small cells~(SCells) deployed, although highly directional antennas and beamforming greatly reduce the co-channel interference, and make it possible to overcome the high near-field path loss and poor diffraction of mmWave signals \cite{rappaport2014millimeter,TWC2019Gao,JSAC2017Yu_antenna,CM2018yang,TVT2019Hang}. A feasible scenario is that mmWave SCells are overlaid on traditional Sub-6GHz networks, where the Sub-6GHz and mmWave base stations~(BSs) provide universal coverage and high data rate transmission in traffic hotspots, respectively \cite{TC2017mmTut_Andrews}. In addition, unlike Sub-6GHz BSs with omnidirectional antennas, providing an initial access for stand alone mmWave BSs is challenging due to highly directional mmWave communications \cite{glocom2016mminitial,twc2017initial}. As such, a promising solution is to deploy Sub-6GHz BSs together with mmWave BSs, which can assist the initial access of mmWave communications through sharing the positions and orientations of BSs.

Lately, there exists several studies concentrating on the integrated Sub-6GHz and mmWave cellular networks. Since stochastic geometry is a unified mathematical paradigm to analyze the performance of cellular networks \cite{CST2017ElSawy_SGtut,andrews2011tractable,kong2016jsacginibre,flint2017twcphchp}, it is likely to model the locations of BSs in each tier as a independent Poisson point process~(PPP). Under this condition, the signal-to-interference-plus-noise (SINR) coverage probability with decoupled cell association strategy was studied in \cite{TWC2016De_Elshaer}, and it is observed that extremely high small cell association weight is desirable for mmWave SCells. The authors in \cite{tcom2018yi} analyzed the the performance of catch-enable hybrid heterogeneous networks under the similar deployment settings, where the cached multimedia contents following the popularity rank, and it is shown that the integrated Sub-6GHz and mmWave HetNet is interference-limited and outperforms the traditional HetNet. 
The hybrid cellular network with ultra high frequency and mmWave BSs were investigated in \cite{icc2016omar} by using experimental data in a university campus. It is confirmed that the hybrid cellular network could achieve better SINR and rate coverage than those of stand alone ultra high frequency network and the mmWave network. 
Besides, a device-to-device~(D2D) communication model with hybrid frequency was investigated in \cite{vtc2016multibandWang}, where the user equipments~(UEs) employ mmWave communication when there is no blockage and switch to Sub-6GHz otherwise. The results also demonstrate the superiority of the hybrid communication model. 

Although PPP is tractable in modeling random networks, it is not rich enough in capturing spatial coupling between UE and BS locations that exists in traffic hotspots \cite{saha2018pcp,saha2018pcp0}, which can be better modeled by Poisson cluster process (PCP). The PCP-based modeling and performance analysis of HetNet has gained much attention in these years \cite{saha2018pcp0,ganti2009pcp,afshang2018}. 
A complete characterization of the downlink coverage probability for a PCP-based HetNet model under max-SINR based association scheme is investigated in \cite{saha2018pcp}. 
The BS-centric cellular network was analyzed in \cite{ppp2016mankar}, in which the locations of BSs are modeled as a PPP, and the UEs are modeled as a PCP around BSs. The results show that the coverage experienced under the PCP model becomes the same as that under the PPP model when the standard deviation of the PCP tends to infinity. As an extension, the BS-centric multi-tier case is further investigated in \cite{kppppcp2016saha}.
In contrast, the user-centric capacity-driven small cell deployment is proposed in \cite{pcp20172tier}, where both the BSs and UEs are modeled as PCPs in user hotspots. It is shown that higher frequency reuse brings lower coverage probability and higher throughput.
Besides, the PCP-based model can be applied to D2D network as \cite{glocom2015d2dpcpAfshang}, where the devices inside a given cluster form D2D links amongst themselves according to the fact that D2D devices need to be in close proximity of each other. 
Moreover, the distance of the PCP distributed BS from the receiver under max-power based association is evaluated using nearest-neighbor and contact distance distributions in \cite{afshang2016nncd}. 
To reduce the numerical complexity, a computable form of coverage probability for the downlink cellular network with PCP distributed BSs, Rayleigh fading and nearest BS association is analyzed in \cite{miyoshi2019pcp}. 

To the authors' knowledge, the existing works that considering integrated Sub-6GHz BSs and mmWave BSs usually model each tier of BSs as a PPP \cite{TWC2016De_Elshaer,tcom2018yi}, while the studies that consider spatial coupling between BSs and UEs in heterogeneous networks assume the operations on the same frequency band \cite{ppp2016mankar,kppppcp2016saha,pcp20172tier}. Different from previous works, we take both different frequency bands and aforementioned coupling in traffic hotspots into consideration. More specifically, the Sub-6GHz and mmWave BSs are modeled as a PPP and a Thomas cluster process (TCP), respectively. The main contributions of this paper are summarized as follows. 

\begin{itemize}
	\item
	We investigate a two-tier heterogeneous network consisting of independently distributed Sub-6GHz macrocells (MCells) and clustered mmWave SCells to improve the network performance in traffic hotspots with small-scale fading modeled as  Nakagami-$m$ fading, where different propagation characteristics of Sub-6GHz and mmWave bands are considered. The UEs cluster around hotspot centers following a TCP, and apply the strongest bias association strategy.
	\item
	We derive the general expressions of association probability, SINR coverage probability, average achievable rate, and area throughput via stochastic geometry-based analysis. Furthermore, we simplify the above expressions by considering the LoS signals alone in mmWave propagation. Moreover, a special case of the two-tier Sub-6GHz network under the same distribution is considered as a baseline scheme, which shows that the clustered mmWave BSs could enhance the coverage substantially.
	\item
	We analyze the impact of various parameters on the network performance both theoretically and numerically. Firstly, we show that the distribution standard deviation, which is used to scale the degree of dispersion of TCP, is an important parameter in determining the coverage probability. And there exists an optimal distribution standard deviation of mmWave BSs that is proportional to the distribution standard deviation of clustered UEs and maximizes the coverage probability. Then, the ratio of association weight for Sub-6GHz and mmWave BSs is shown to be a prominent factor and needs to be set properly. In particular, extremely high and extremely small association weight for mmWave BSs will deteriorate the performance for cell edge users and cell interior users, respectively. Finally, several different deployment schemes in traffic hotspots are investigated, revealing that the UEs far from hotspot centers indeed need Sub-6GHz service to achieve an acceptable SINR coverage.
\end{itemize}

The rest of the paper is organized as follows. The system model is presented in Section~\ref{sec:model}. In Section~\ref{sec:analysis}, the expressions of association probability, SINR coverage probability, average achievable rate, and area throughput are derived. In Section~\ref{sec:simulation}, the numerical results are presented, and the impacts of various parameters on the network performance are also investigated. The conclusions are drawn in Section~\ref{sec:conclusion}.

\section{System Model}\label{sec:model}
In this section, we first provide a brief introduction to TCP before we introduce the proposed system model.

TCP is a stationary and isotropic Poisson cluster process generated by a set of parent points independently and identically distributed around each point of a parent PPP \cite{ganti2009pcp}. In particular, the locations of parent points are modeled as a homogeneous PPP $\Phi$ with density $\lambda$. For each parent point $\bm{c}\in\Phi$, the daughter points are scattered following a symmetric normal distribution with variance $\sigma^2$. The probability density function (PDF) of an daughter point location relative to its parent point can be expressed as $f\left(\bm{x}\right) = \frac{1}{2\pi\sigma^2} \exp\left( -\frac{\lVert\bm{x}-\bm{c}\rVert^2}{2\sigma^2} \right)$. In addition, the number of daughter points in each cluster is a Poisson random variable with mean $n$. Therefore, the TCP can be characterized by its parent points set $\Phi$, the distribution standard deviation $\sigma$ and the cluster size $n$. For simplicity of notation, we denote the aforementioned TCP by $\mathcal{G}\left(\Phi,\sigma,n\right)$.

\begin{figure}
	\centering
	\includegraphics[width=4in]{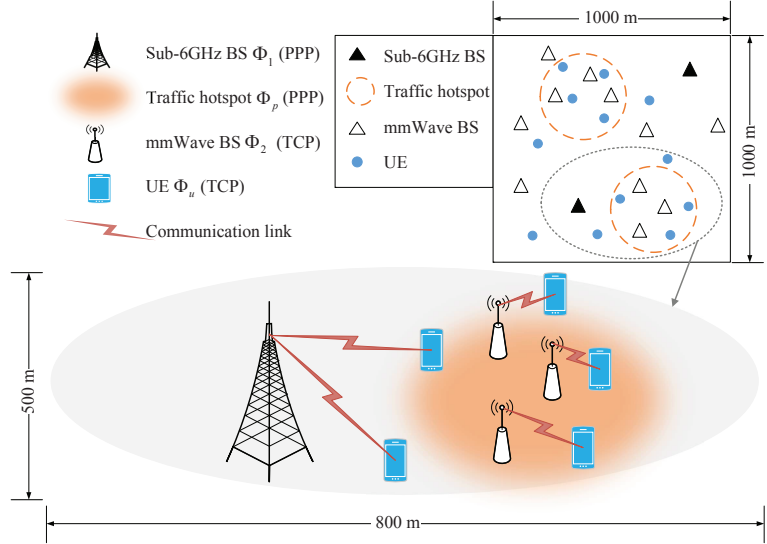}
    \caption{Layout of the proposed system model with PPP distributed Sub-6GHz BSs and TCP distributed mmWave BSs in $1~\text{km} \times 1~\text{km}$ area, $\lambda_1=2~/\text{km}^2$, $\lambda_p=2~/\text{km}^2$ and $n_\bs=5$. The mmWave BSs are clustered around hotspot centers following symmetric normal distribution with standard deviation $\sigma=200$.}
	\label{model}
\end{figure}

\subsection{Spatial Model}\label{subsec:spatial}
We consider a downlink two-tier cellular network with hotspots as shown in Fig.~\ref{model}, where the hotspot centers $\Phi_p=\left\{\bm{c}_0,\bm{c}_1,\ldots\right\}$ is a homogeneous PPP with density $\lambda_p$, and $\bm{c}_i$ represents the location of the hotspot center with index $i$. The first tier of BSs $\Phi_1$ is assumed to be operated at Sub-6GHz and modeled as a homogeneous PPP with density $\lambda_1$, while the second tier is operated at mmWave and distributed as a TCP $\mathcal{G}\left(\Phi_p,\sigma_\bs,n_\bs\right)$. The set of mmWave BSs generated by $\bm{c}_i$ is denoted by $\mathcal{X}_{\bm{c}_i}$. In addition, to capture the relativity between UEs and traffic hotspots, UEs $\Phi_u$ are assumed to follow another TCP with parent points $\Phi_p$, i.e., $\Phi_u=\mathcal{G}\left(\Phi_p,\sigma_\ue,n_\ue\right)$. 

The transmit powers of Sub-6GHz BSs and mmWave BSs are set to be $P_1$ and $P_2$, respectively. Without loss of generality, our analysis is conducted on the typical UE $\bm{y}_0$, which is generated from the parent point $\bm{c}_0$ and is located at the origin. Moreover, to facilitate the analysis and maintain tractability, we assume that the number of mmWave SCells in the typical UE belonged cluster is constant and equal to $n_\bs$.

\subsection{Directional Beamforming}\label{subsec:antenna}
The mmWave BSs are equipped with directional antenna array to improve beamforming gains and to compensate for the high path loss. However, we only consider an omnidirectional antenna model at Sub-6GHz BSs and UEs for tractability of analysis. The omnidirectional antenna gain at Sub-6GHz BSs side is $G_1$, and directional antenna arrays of mmWave BSs are approximated by the sectored antenna model \cite{Bai2014CM_mm}, namely
\begin{align}\label{eq:beamforming}
G_\mathsf{b}\left(\theta\right)=
	\begin{cases}
	G_\mathsf{M},&\text{if $\lvert\theta\rvert \leq {\theta_\mathsf{b}}/2$},
	\\
	G_\mathsf{m},&\text{otherwise},
	\end{cases}
\end{align}
where $\theta_\mathsf{b}$ is the beamwidth of the main lobe, and $G_\mathsf{M}$ and $G_\mathsf{m}$ are the main-lobe and side-lobe gains, respectively. When the typical UE is associated with an mmWave BS, the mmWave BS first estimates the channel, and then adjusts its antenna steering orientation to the typical UE to maximize the directivity gain $G_\mathsf{b}\left(\theta\right)$. Due to the isotropy of Thomas cluster process, the beam directions of the interference links are independently and uniformly distributed in $\left[-\pi,\pi\right]$. Therefore, the antenna gain of a randomly chosen interfering mmWave BS is $G_\mathsf{M}$ with probability $p_\mathsf{M} = \theta_\mathsf{b}/\left(2\pi\right)$, and is $G_\mathsf{m}$ with probability $p_\mathsf{m} = 1-\theta_\mathsf{b}/\left(2\pi\right)$.

\subsection{Channel Model}\label{subsec:channel}
A communication link is either line-of-sight~(LoS) or non-line-of-sight~(NLoS), depending on whether the BS is visible to the typical UE or not. In Sub-6GHz networks, the links are usually long and NLoS, which is already considered in the path loss exponent. However, for mmWave networks, the links usually work with shorter distance and are more sensitive to the blockage effects, and thus different path loss exponents are needed to model LoS/NLoS mmWave links. Here, we adopt the generalized blockage model to characterize the mmWave propagations \cite{JSAC2015Singh_backhaul}, i.e., the probability function of LoS follows $P_\mathsf{L}\left(r\right) = p_\mathsf{L} \cdot \mathbf{1}\left( r < R_\mathsf{B} \right)$, where $\mathbf{1}\left(\cdot\right)$ is the indicator function, $r$ is the distance between the mmWave BS and the typical UE, $R_\mathsf{B}$ is the size of the LoS ball, and $p_\mathsf{L}$ is the average fraction of the LoS area in the LoS ball.

Different path loss intercepts and exponents shall be adopted for the signals on different frequency bands with LoS/NLoS status. Given a communication link with length $r$ in the investigated two-tier cellular networks, the path loss is formulated as $\ell_k\left(r\right) = C_k r^{-\alpha_k}$, where $\alpha_k$ is the path loss exponent with $k\in\left\{1,\mathsf{L},\mathsf{N}\right\}$, and $C_k$ denotes the free space path loss at $1$~m with carrier frequency $f_k$. Here, the indices of ``1'', ``$\mathsf{L}$'' and ``$\mathsf{N}$'' correspond to the cases of Sub-6GHz, mmWave LoS and mmWave NLoS, respectively.

Furthermore, we assume independent Nakagami-$m$ fading with integer parameter $N_k$ for each link. Let $h_{\bm{x}}$ denote the small-scale fading gain of the link $\bm{x}\rightarrow\bm{y}_0$,  $h_{\bm{x}}$ is a normalized Gamma random variable with distribution $\Gamma\big( N_k,\frac{1}{N_k} \big)$. Since the small-scale fading for Sub-6GHz band is predicated on a large amount of local scattering \cite{TC2017mmTut_Andrews}, we assume Rayleigh fading for Sub-6GHz propagations, i.e., $N_1=1$.
Besides, we ignore the large scale shadowing effect as in \cite{TWC2016De_Elshaer}, since the blockage model for mmWaves introduces a similar effect to shadowing \cite{TC2017mmTut_Andrews}, and the randomness of the PPP distributed Sub-6GHz BS locations emulates the shadowing effect \cite{CST2017ElSawy_SGtut}.

\subsection{Association Strategy}\label{subsec:strategy}
Each UE is associated with the BS with the maximum bias averaged received power, namely the strongest bias association strategy. Therefore, the serving BS of the typical UE is expressed as
\begin{align}\label{eq:associationStrategy}
	\bm{x}^*
	= \mathop{\arg\max}_{\bm{x} \in \Phi_1\cup\Phi_2}
	B_k P_k G_k N_k \ell_k\left(\lVert \bm{x} \rVert\right),
	\quad k\in\left\{ 1,2 \right\},
\end{align}
where $B_k$ is the bias association value of the $k$th tier, $\lVert\bm{x}\rVert$ is the distance from BS $\bm{x}$ to the typical UE, $P_k$ and $G_k$ are the transmit power and the maximum antenna gain of the BS in the $k$th tier, respectively, and $k=2$ represents the case of mmWave association for notational simplicity. It is worth noting that the parameter $B_k$, also known as cell range expansion parameter \cite{TWC2012HetNet_Jo}, is able to offload users between different tiers. Moreover, to guarantee the quality of service, the serving BS is confined to be Sub-6GHz or mmWave LoS, i.e., mmWave NLoS BSs are neglected due to the high path loss. Note that the traffic hotspots are sparsely deployed, which means that the distance between the typical UE and mmWave BSs in different traffic hotspots are usually long, and we further assume the potential mmWave serving BSs to be intra-cluster mmWave LoS BSs. Note that this assumption is just for simplifying the analysis, and the results can be extended to the general association policy following similar analysis methods.

When the typical UE is associated with the $k$th tier, $k\in\left\{ 1,2 \right\}$, the downlink received SINR can be expressed as
\begin{align}\label{eq:SINRexpression}
	\sinr_k
	= \frac{P_k G_k h_{\bm{x}^*} \ell_k\left(\lVert \bm{x}^* \rVert\right)}{\sigma_k^2 + I_k},
\end{align}
where $\sigma_k^2$ is the thermal noise, and $I_k$ is the aggregate interference. For $k=1$, $I_1$ is expressed as
\begin{align}\label{eq:interfer1Expression}
	I_1
	= \sum_{\bm{x}\in\Phi_1\backslash\bm{x}^*}
	P_1 G_1 h_{\bm{x}} \ell_1\left(\lVert \bm{x} \rVert\right).
\end{align}
For $k=2$, $I_2$ can be separated into intra-cluster interference $I_2^\mathsf{intra}$ and inter-cluster interference $I_2^\mathsf{inter}$, and is expressed as
\begin{align}\label{eq:interfer2Expression}
	I_2
	= \underbrace{\sum_{\bm{x}\in\mathcal{X}_{\bm{c}_0}\backslash\bm{x}^*}
	P_2 G_\mathsf{b}\left(\theta\right) h_{\bm{x}} \ell_2\left( \lVert\bm{x}\rVert \right)}_{ I_2^\mathsf{intra} }
	+ \underbrace{\sum_{\bm{c}_i\in\Phi_p\!\backslash\bm{c}_0} \sum_{\bm{x}\in\mathcal{X}_{\bm{c}_i}}
	P_2 G_\mathsf{b}\left(\theta\right) h_{\bm{x}} \ell_2\left( \lVert\bm{x}\rVert \right)}_{ I_2^\mathsf{inter} }.
\end{align}

It can be seen that the SINR in \eqref{eq:SINRexpression} is a random variable due to the randomness of BS locations, UE locations, antenna gain and small scale fading. Using the tools of stochastic geometry, we can evaluate the performance of coverage probability and throughput in the following section.

\section{Network Performance Analysis}\label{sec:analysis}
To investigate the network performance, we begin by deriving several auxiliary results on distance distributions and the probability of the typical UE being associated with each tier in Section~\ref{subsec:AssociationAnalysis}. And then we derive the expressions of SINR coverage probability in Section~\ref{subsec:SINRCoverageAnalysis}, and the results of average achievable rate in Section~\ref{subsec:RateThroughputAnalysis}. A summary of all the derived lemmas, theorems and corollaries is given in Section~\ref{subsec:logic_flow} to make the logic flow clear.

\subsection{Association Analysis}\label{subsec:AssociationAnalysis}
According to the strongest bias association strategy, the potential serving BSs for the typical UE $\bm{y}_0$ can be the nearest Sub-6GHz BS $\bm{x}^\ast_1$ in $\mathbb{R}^2$ or the nearest mmWave LoS BS $\bm{x}^\ast_2$ in $\mathcal{X}_{\bm{c}_0}$. Let $R_k$ denote $\lVert \bm{y}_0-\bm{x}^\ast_k\rVert$, $k\in\left\{1,2\right\}$, the following Lemma provides the distribution of $R_k$.

\begin{lemma}\label{lemma:minDistancePDF}
The cumulative distribution function~(CDF) and PDF of $R_k$, conditioned on $\lVert \bm{y}_0 - \mathbf{c}_0 \rVert = v_0$, are given by
\begin{align}\label{eq:F_R}
	F_{R_k}\left(r;v_0\right) =
	\begin{cases}
		1 - \exp\left(-\pi\lambda_1 r^2\right), &k=1,
		\\
		1 - \left[ 1-F_{S_\mathsf{L}}\left(r;v_0\right) \right]^{n_\bs}, &k=2,
	\end{cases}
\end{align}
and
\begin{align}\label{eq:f_R}
	f_{R_k}\left(r;v_0\right) =
	\begin{cases}
		2\pi\lambda_1 r\exp\left(-\pi\lambda_1 r^2\right), &k=1,
		\\
		n_\bs \left[
		1-F_{S_\mathsf{L}}\left(r;v_0\right) \right]^{n_\bs-1}
		f_{S_\mathsf{L}}\left(r;v_0\right), &k=2,
	\end{cases}
\end{align}
respectively, where $S_\mathsf{L}$ is the distance from the typical UE to a randomly chosen mmWave LoS BS in $\mathcal{X}_{\bm{c}_0}$. The CDF and PDF of $S_\mathsf{L}$ are given by
\begin{align}
	F_{S_\mathsf{L}}\left(r;v_0\right)
	&= \int_0^r \frac{t}{2\pi\sigma_\bs^2}
	\exp\left( -\frac{t^2+v_0^2}{2\sigma_\bs^2} \right)
	P_\mathsf{L}\left( t \right)
	J\left( \frac{v_0 t}{\sigma_\bs^2} \right)
	\,\mathrm{d}t,
\end{align}
and
\begin{align} \label{eq:PDF:SL}
	f_{S_\mathsf{L}}\left(r;v_0\right)
	&= \frac{P_\mathsf{L}\left(r\right) r}{2\pi\sigma_\bs^2}
	\exp{\left( -\frac{r^2+v_0^2}{2\sigma_\bs^2} \right)}
	J\left( \frac{v_0 r}{\sigma_\bs^2}\right),
\end{align}
respectively, where $J\left(t\right) = \int_{-\pi}^{\pi} e^{t\cos\theta}\,\mathrm{d}\theta$.
\end{lemma}

\begin{IEEEproof}
See Appendix \ref{appendix:minDistancePDF}.
\end{IEEEproof}

Let $K$ denote the index of the tier that the typical UE is associated with, and let $\mathcal{A}_k^\mathsf{c}\left(v_0\right)$ denote the probability of $K=k$ in the presence of $\lVert \bm{y}_0 - \mathbf{c}_0 \rVert = v_0$, namely the conditional association probability of the $k$th tier. The following lemma provides the expression of $\mathcal{A}_k^\mathsf{c}\left(v_0\right)$.

\begin{lemma}\label{lemma:A_c}
The conditional association probability $\mathcal{A}_k^\mathsf{c}\left(v_0\right)$ is given by
\begin{align}
	\mathcal{A}_1^\mathsf{c}\left(v_0\right)
	&= \int_0^\infty
	2\pi\lambda_1 r\exp\left(-\pi\lambda_1 r^2\right)
	\bar{F}_{S_\mathsf{L}}\left[ \delta_{1,2}^{n_\bs} \left( r \right); v_0 \right]
	\,\mathrm{d}r, \label{eq:A_c_1}
	\\
	\mathcal{A}_2^\mathsf{c}\left(v_0\right)
	&= \int_0^\infty
	n_\bs \bar{F}_{S_\mathsf{L}}\left(r;v_0\right) ^{n_\bs-1}  f_{S_\mathsf{L}}\left(r;v_0\right)
	\exp\left[-\pi\lambda_1  \delta_{2,1}^2 \left( r \right) \right]
	\,\mathrm{d}r, \label{eq:A_c_2}
\end{align}
where $\delta_{1,2}\left( r \right) = \left(\frac{B_2 P_2 G_\mathsf{M} N_\mathsf{L}}{B_1 P_1 G_1 N_1}\right)^\frac{1}{\alpha_\mathsf{L}} r^\frac{\alpha_1}{\alpha_\mathsf{L}}$ and $\delta_{2,1}\left( r \right) = \left(\frac{B_1 P_1 G_1 N_1}{B_2 P_2 G_\mathsf{M} N_\mathsf{L}}\right)^\frac{1}{\alpha_1} r^\frac{\alpha_\mathsf{L}}{\alpha_1}$.
\end{lemma}

\begin{IEEEproof}
Conditioned on $\lVert \bm{y}_0 - \mathbf{c}_0 \rVert = v_0$, the typical UE is associated with Sub-6GHz BS if and only if $B_1 P_1 G_1 N_1 \ell\left(R_1\right) > B_2 P_2 G_\mathsf{M} N_2 \ell\left(R_2\right)$. Thus $\mathcal{A}_1^\mathsf{c}\left(v_0\right)$ can be formulated as
\begin{align}\label{proof_lemma2}
    \mathcal{A}_1^\mathsf{c}\left(v_0\right)
    &= \mathbb{P}\left[ B_1 P_1 G_1 N_1 \ell\left(R_1\right) > B_2 P_2 G_\mathsf{M} N_\mathsf{L} \ell\left(R_2\right) \right] \\
    &= \mathbb{P}\left[
    R_2 > \delta_{1,2}\left( R_1 \right) \right] \\
    &= \int_0^\infty f_{R_1}\left(r;v_0\right)
	\bar{F}_{R_2}\left[ \delta_{1,2}\left( r \right) \right]
	\, \mathrm{d}r,
\end{align}
where $\delta_{1,2}\left( r \right) = \left(\frac{B_2 P_2 G_\mathsf{M} N_\mathsf{L}}{B_1 P_1 G_1 N_1}\right)^\frac{1}{\alpha_\mathsf{L}} r^\frac{\alpha_1}{\alpha_\mathsf{L}}$. And $\mathcal{A}_2^\mathsf{c}\left(v_0\right)$ can be calculated following on the same lines.
\end{IEEEproof}

Since the UEs are distributed in traffic hotspots following independent Gaussian distribution with standard deviation $\sigma_\ue$, the performance experienced by the typical UE depends on its distance to the hotspot center. Based on Lemma~\ref{lemma:A_c}, we can derive the association probability that averaged over the distance to the hotspot center, as shown in the following theorem.

\begin{theorem}\label{theorem:AsPr}
The probability that the typical UE is associated with the $k$th tier is
\begin{align}\label{eq:Ak}
	\mathcal{A}_k
	= \int_0^\infty
	\frac{v_0}{\sigma_\ue^2}
	\exp\left( -\frac{v_0^2}{2\sigma_\ue^2} \right)
	\mathcal{A}_k^\mathsf{c}\left(v_0\right)
	\,\mathrm{d}v_0.
\end{align}
\end{theorem}

\begin{IEEEproof}
The expression of $f_{V_0}\left(v_0\right)$ can be derived by using polar coordination
\begin{align}\label{eq:f_V0}
	f_{V_0}\left(v_0\right)
	= \int_0^{2\pi}
	f_{\bm{Y}}\left( v_0,\theta \right) v_0
	\,\mathrm{d}\theta
	= \frac{v_0}{\sigma_\ue^2}
	\exp\left( -\frac{v_0^2}{2\sigma_\ue^2} \right),
\end{align}
where $f_{\bm{Y}}\left( v_0,\theta \right)$ is the PDF of UEs with respect to their cluster centers. The proof can be obtained by using the fact that $\mathcal{A}_k=\mathbb{E}_{V_0}\left[ \mathcal{A}_k^\mathsf{c}\left(v_0\right) \right]$, and taking expectation with respect to $V_0$.
\end{IEEEproof}

Let $X_k$ denote the conditional distance $\left\{ \left.X_k=R_k\right|K=k, V_0=v_0 \right\}$, its PDF can be derived from Lemma~\ref{lemma:A_c} and Theorem~\ref{theorem:AsPr} as follows.

\begin{corollary}\label{corollary:Xk}
The PDF of conditional distance $X_k$ is given by
\begin{align}\label{eq:X_k}
	f_{X_1}\left(x;v_0\right)
	&= \frac{2\pi\lambda_1 x}{\mathcal{A}_1^\mathsf{c}\left(v_0\right)}
	\exp\left(-\pi\lambda_1 x^2\right)
	\bar{F}_{S_\mathsf{L}}\left[ \delta_{1,2}^{n_\bs} \left( x \right); v_0 \right],
	\\
	f_{X_2}\left(x;v_0\right)
	&= \frac{n_\bs}{\mathcal{A}_2^\mathsf{c}\left(v_0\right)}
	\bar{F}_{S_\mathsf{L}}\left(x;v_0\right) ^{n_\bs-1}  f_{S_\mathsf{L}}\left(x;v_0\right)
	\exp\left[-\pi\lambda_1  \delta_{2,1}^2 \left( x \right) \right].
\end{align}
\end{corollary}

\begin{IEEEproof}
Leveraging the conditional probability formula, we have
\begin{align}\label{proof_corollary1}
	F_{X_k}\left(x;v_0\right)
	&= \mathbb{P}\left( \left.R_k\leq x \right| K=k, V_0=v_0 \right) \\
	&= \frac{\mathbb{P}\left( \left.R_k \leq x, K=k \right| V_0=v_0 \right)}
    {\mathbb{P}\left( \left.K=k \right| V_0=v_0 \right)} \\
    &= \frac{1}{\mathcal{A}_k^\mathsf{c}\left(v_0\right)}
    \mathbb{P}\left[ \left.R_k \leq x, R_{3-k} > \delta_{k,3-k}\left( R_k \right) \right| V_0=v_0 \right] \\
	&= \frac{1}{\mathcal{A}_k^\mathsf{c}\left(v_0\right)}
	\int_0^x f_{R_k}\left(r\right)
	\bar{F}_{R_k}\left[ \delta_{k,3-k}\left( x \right);v_0 \right]
	\, \mathrm{d}r.
\end{align}
By taking the derivative of $F_{X_k}\left(x;v_0\right)$ with respect to $x$, we obtain
\begin{align}\label{proof_fx}
    f_{X_k}\left(x;v_0\right)
    &= \frac{1}{\mathcal{A}_k^\mathsf{c}\left(v_0\right)}
	f_{R_k}\left(x\right)
	\bar{F}_{R_k}\left[ \delta_{k,3-k}\left( x;v_0 \right) \right].
\end{align}
\end{IEEEproof}

\subsection{SINR Coverage Analysis}\label{subsec:SINRCoverageAnalysis}
Here, we provide the general expression of the SINR coverage probability $\mathcal{C}\left(\tau\right)$, which is defined as the probability that the instantaneous received SINR is greater than a threshold $\tau$, i.e., $\mathbb{P}\left[\sinr>\tau\right]$. Although the mmWave cellular networks are usually considered to be noise-limited \cite{TC2017mmTut_Andrews,CM2017an}, the mmWave interference in our model is non-negligible due to the high local density in traffic hotspots, which is shown in Section~\ref{sec:simulation}.

Based on the results in Section~\ref{subsec:AssociationAnalysis}, the SINR coverage probability can be evaluated as
\begin{align}\label{CovTauSplit}
	\mathcal{C}\left(\tau\right)
	= \mathbb{E}_{V_0}\big[
	\mathcal{C}\left(\tau;v_0\right)
	\big]
	= \mathbb{E}_{V_0}\left[
	\sum_{k=1}^2
	\mathcal{A}_k^\mathsf{c}\left(v_0\right) \mathcal{C}_k\left(\tau;v_0\right)
	\right],
\end{align}
where $\mathcal{C}_k\left(\tau;v_0\right)$ is the SINR coverage probability in the presence of $K=k$ and $V_0=v_0$. Since the distinguishing features of Sub-6GHz and mmWave are incorporated, we analyze the cases of Sub-6GHz association and mmWave association in sequence as follows.

In the presence of Sub-6GHz association, i.e., $k=1$, the aggregate interference $I_1$ is received from all the other Sub-6GHz BSs in the plane. By leveraging the properties of PPP, the conditional SINR coverage probability $\mathcal{C}_1\left( \tau;v_0 \right)$ can be evaluated in Lemma~\ref{lemma:C_1_con}.

\begin{lemma}\label{lemma:C_1_con}
$\mathcal{C}_1\left( \tau;v_0 \right)$ is given by
\begin{align}\label{eq:SINRcov1}
	\mathcal{C}_1\left( \tau;v_0 \right)
	&= \int_0^\infty
	f_{X_1}\left( x; v_0 \right)
	\exp\left( -\frac{x^\alpha\tau\sigma_1^2}{b_1} \right)
	\LT{I_1}{\frac{x^\alpha\tau}{b_1};v_0,x}
	\,\mathrm{d}x,
\end{align}
where
\begin{align}\label{eq:LT1}
	\LT{I_1}{s;v_0,x}
	= \exp\left\{ - 2\pi\lambda_1
	\int_x^\infty
	\left[ 1-\left(1+ s b_1 r^{-\alpha}\right)^{-1} \right] r
	\,\mathrm{d}r \right\},
\end{align}
and $b_1=P_1 G_1 C_1$.
\end{lemma}
\begin{IEEEproof}
See Appendix \ref{appendix:C_1_con}.
\end{IEEEproof}

In the presence of mmWave association, i.e., $k=2$, the aggregate interference $I_2$ can be separated into intra-cluster interference $I_2^\intra$ and inter-cluster interference $I_2^\inter$. Since Nakagami-$m$ fading is assumed for mmWave communications, the extract coverage results cannot be obtained analytically. Here we derive an approximate expression of $\mathcal{C}_2\left( \tau;v_0 \right)$ in Lemma~\ref{lemma:C_2_con} by using Alzer's inequality \cite{alzer1997some}. This approximation method has been shown to be generally tight in numerical simulations with different system parameters \cite{TC2017mmTut_Andrews}.

Moreover, note that under the assumptions in Section~\ref{sec:model}, the number of daughter points of each TCP cluster is a Poisson random variable, and thus the mmWave BSs of some traffic hotspots might be empty, which will reduce the tractability in our analysis. Consequently, the derivation of Lemma~\ref{lemma:C_2_con} will be conducted on existence of a mmWave BS for each cluster. 

\begin{lemma}\label{lemma:C_2_con}
$\mathcal{C}_2\left( \tau;v_0 \right)$ can be evaluated as
\begin{align}\label{eq:SINRcov2}
	&\mathcal{C}_2\left( \tau;v_0 \right)
	\approx \int_0^\infty f_{X_2}\left( x;v_0 \right) \left[
	\sum_{n=1}^{N_\mathsf{L}}
	\left(-1\right)^{n+1}
	\binom{N_\mathsf{L}}{n} \right.\nonumber
	\\
	&\qquad\qquad\qquad\left.\times\exp\left( -\frac{ x^{\alpha_\mathsf{L}} \tau\sigma_2^2 \chi_2 n }{b_2} \right)
	\LT{I_2}{ \frac{ x^{\alpha_\mathsf{L}} \tau \chi_2 n }{b_2}; v_0, x }
	\right] \,\mathrm{d}x,
\end{align}
where
\begin{align}
	&\LT{I_2}{s;v_0,x}
	= \LT{I_2^\intra}{s;v_0,x}
	\cdot \LT{I_2^\inter}{s;v_0,0}, \label{eq:LT2}
	\\
	&\LT{I_2^\intra}{s;v_0,x}
	= \exp\Bigg\{ -2\pi\left(n_\bs-1\right)
	\sum_{ i\in\left\{\mathsf{L,N}\right\} }
	\int_{ \delta_{\mathsf{L},i}\left(x\right) }^\infty
	f_{X_2}\left(r;v_0\right) \nonumber
	\\
	&\qquad\qquad\qquad\qquad\qquad \times\sum_{ \mathclap{j\in\left\{\mathsf{m,M}\right\}} } p_j \left[
	1 - \left(
	1 + s b_{2,j} r^{-\alpha_\mathsf{L}} \right)^{-N_\mathsf{L}}
	\right] r
	\,\mathrm{d}r \Bigg\}, \label{eq:LT2_intra}
	\\
	&\LT{I_2^\inter}{s;v_0,0}
	= \exp\left\{ -2\pi\lambda_p
	\int_0^\infty
	\left[ 1 - \LT{\left.I_\mathsf{L}^\intra\right|n_\bs+1}{s;v,0} \right] v
	\,\mathrm{d}v \right\}, \label{eq:LT2_inter}
\end{align}
with $\chi_2 = N_\mathsf{L} \left( N_\mathsf{L}! \right)^{-\frac{1}{N_\mathsf{L}}}$, $b_2=P_2 G_\mathsf{M} C_\mathsf{L}$,
and $b_{2,j}=P_2 G_j C_\mathsf{L}$, $j\in\left\{ \mathsf{m},\mathsf{M} \right\}$.
\end{lemma}

\begin{IEEEproof}
See Appendix \ref{appendix:C_L_con}.
\end{IEEEproof}

Employing the substitution of Lemma~\ref{lemma:C_1_con} and Lemma~\ref{lemma:C_2_con} in \eqref{CovTauSplit}, we can derive the SINR coverage probability of the network, as shown in Theorem \ref{theorem:SINRcov}.
\begin{theorem}\label{theorem:SINRcov}
The SINR coverage probability $\mathcal{C}\left(\tau\right)$ is given by
\begin{align}\label{eq:overallCoverage}
	\mathcal{C}\left(\tau\right)
	\approx& \int_0^\infty
	f_{V_0}\left( v_0 \right) \left[
	\sum_{k=1}^2
	\mathcal{A}_k^\mathsf{c}\left(v_0\right)
	\int_0^\infty f_{X_k}\left( x;v_0 \right)
	\right.\nonumber
	\\
	&\quad\times\left.
	\sum_{n=1}^{N_k} a_k\left(n\right)
	\exp\left( -\frac{x^{\alpha_k} \tau\sigma_k^2 \chi_k n}{b_k} \right)
	\LT{I_k}{ \frac{x^{\alpha_k} \tau \chi_k n}{b_k}; v_0, x }
	\,\mathrm{d}x \right]
	\,\mathrm{d}v_0
\end{align}
where $a_1\left(n\right)=1$, $a_2\left(n\right)=\left(-1\right)^{n+1} \binom{N_\mathsf{L}}{n}$, and $\LT{I_k}{s;v_0,x}$ is given by \eqref{eq:LT1} and \eqref{eq:LT2}.
\end{theorem}

\begin{IEEEproof}
The proof is obtained by taking the expectation of $\mathcal{C}\left(\tau;v_0\right)$ with respect to $V_0$ and evaluating the integral.
\end{IEEEproof}

Now, we consider special cases, where the expressions of SINR coverage probability can be simplified. These special cases are, respectively, performed on the following assumptions:~\textit{1)} the interference of mmWave NLoS links are neglected,~\textit{2)} the two-tier network is performed on Sub-6GHz band, i.e., replacing mmWave SCells with Sub-6GHz SCells.

\subsubsection{No mmWave NLoS Interference}
Since the mmWave NLoS links are blocked by buildings and suffer from high path loss, the mmWave NLoS interference is marginal in contrast with LoS interference, and we neglect the mmWave NLoS interference. The expression of SINR coverage probability in this case is given by Corollary \ref{corollary:2:Case1}.

\begin{corollary}\label{corollary:2:Case1}
If the mmWave NLoS interference with respect to the typical UE is neglected, the SINR coverage probability can be evaluated as \eqref{eq:overallCoverage}, with the Laplace transform of interference $\LT{I_k}{s;v_0,x}$ being formulated as following
\begin{align}
	\LT{I_1}{s;v_0,x}
	&= \exp\left\{ -2\pi\lambda_1
	\int_x^\infty
	\frac{r}{1 + s^{-1}b_1 r^\alpha}
	\,\mathrm{d}r \right\},
	\\
	\LT{I_2}{s;v_0,x}
	&= \LT{I_\mathsf{L}^\intra}{s;v_0,x}
	\cdot\LT{I_\mathsf{L}^\inter}{s;v_0,0},
\end{align}
where
\begin{align}
	&\LT{I_\mathsf{L}^\intra}{s;v_0,x}
	= \exp\left\{ -2\pi\left( n_\bs-1 \right)
	\int_x^\infty
	f_{S_\mathsf{L}}\left(r;v_0\right)
	\sum_{\mathclap{ j\in\left\{\mathsf{m},\mathsf{M}\right\} }} \
	\frac{p_j r}{1+ s^{-1} b_{2,j} r^{\alpha_\mathsf{L}} }
	\,\mathrm{d}r \right\},
	\\
	&\LT{I_\mathsf{L}^\inter}{s;v_0,x}
	= \exp\left\{ -2\pi\lambda_p
	\int_0^\infty
	\left[ 1 - \LT{\left.I_\mathsf{L}^\intra\right|n_\bs+1}{s;v,0} \right] v
	\,\mathrm{d}v \right\}.
\end{align}
\end{corollary}

\begin{IEEEproof}
The proof can be obtained by removing the NLoS interference terms in \eqref{eq:LT2_intra} and \eqref{eq:LT2_inter}.
\end{IEEEproof}

Corollary~\ref{corollary:2:Case1} gives a simple approximate expression of coverage probability. Due to the contribution of neglecting interfering mmWave BSs with NLoS propagations, it is easy to see that $\mathcal{C}^{(1)}\left(\tau\right)$ is an upper bound of $\mathcal{C}\left(\tau\right)$, which is validated to be generally tight in Section~\ref{subsec:sinrCoverageProbability}. This approximation is reasonable for urban areas where the blockage effects are tremendous and mmWave NLoS signals suffer from severe penetration loss.

\subsubsection{Two-tier Sub-6GHz Network}\label{subsec:3_sp2}
To compare the integrated Sub-6GHz-mmWave model with traditional Sub-6GHz model, we consider a baseline two-tier Sub-6GHz cellular network, where the mmWave SCells are replaced with the Sub-6GHz SCells, and investigate the SINR coverage probability $\mathcal{C}^{(2)}\left(\tau\right)$ under the assumption that the Sub-6GHz SCells are equipped with omnidirectional antennas and experience Rayleigh fading with a unit mean. The intra-cell interference is ignored due to the orthogonal multiple access within a cell, and the typical UE receives interference from three parts: MCells interference $I_1^\mathsf{0}$, intra-cluster SCells interference $I_2^\intra$ and inter-cluster SCells interference $I_2^\inter$. In such a scenario, the expression of $\mathcal{C}^{(2)}\left(\tau\right)$ is given in the following corollary.

\begin{corollary}\label{corollary:3:Case2}
Under the same deployment as that presented in Section~\ref{sec:model}, and substituting mmWave SCells with Sub-6GHz SCells, the SINR coverage probability $\mathcal{C}^{(2)}\left(\tau\right)$ can be evaluated as
\begin{align}\label{eq:specialcase2}
	&\mathcal{C}^{(2)}\left(\tau\right)
	= \int_0^\infty
	f_{V_0}\left( v_0 \right) \left[
	\sum_{k=1}^2
	\mathcal{A}_k^\mathsf{c}\left(v_0\right)
	\int_0^\infty f_{X_k}\left( x;v_0 \right)
	\right.\nonumber
	\\
	&\qquad\qquad\qquad \left.\times
	\exp\left( -\frac{x^{\alpha} \tau\sigma_1}{b_k} \right)
	\LT{I_k}{ \frac{x^{\alpha} \tau}{b_k}; v_0, x }
	\,\mathrm{d}x \right]
	\,\mathrm{d}v_0,
\end{align}
where
\begin{align}
	\LT{I_1}{s;v_0,x}
	=& \LT{I_1^0}{s;v_0,x}
	\cdot \LT{I_2^\intra}{ s;v_0,\delta_{1,2}\left(x\right) }
	\cdot\LT{I_2^\inter}{ s;v_0,\delta_{1,2}\left(0\right) },
	\\
	\LT{I_2}{s;v_0,x}
	=& \LT{I_1^0}{ s;v_0,\delta_{2,1}\left(x\right) }
	\cdot \LT{I_2^\intra}{s;v_0,x}
	\cdot\LT{I_2^\inter}{s;v_0,0},
\end{align}
with
\begin{align}
	\LT{I_1^0}{s;v_0,x}
	&= \exp\left\{ -2\pi\lambda_1
	\int_x^\infty
	\frac{r}{1+ s^{-1}b_1 r^{\alpha}}\,\mathrm{d}r \right\},
	\\
	\LT{I_2^\intra}{s;v_0,x}
	&= \exp\left\{ -2\pi\left( n_\bs-1 \right)
	\int_x^\infty
	f_{S_\mathsf{L}}\left(r;v_0\right)
	\frac{r}{1+ s^{-1} b_{2} r^{\alpha} }
	\,\mathrm{d}r \right\},
	\\
	\LT{I_2^\inter}{s;v_0,x}
	&= \exp\left\{ -2\pi\lambda_p
	\int_0^\infty
	\left[ 1 - \LT{\left.I_2^\intra\right|n_\bs+1}{s;v,0} \right] v
	\,\mathrm{d}v \right\}.
\end{align}
\end{corollary}

\begin{IEEEproof}
The proof is obtained by applying substitutions $P_\mathsf{L}\left(r\right) \rightarrow 1$, $G_\mathsf{M} \rightarrow G_1$, $G_\mathsf{m} \rightarrow G_1$, $\ell_2\left(r\right) \rightarrow \ell_1\left(r\right)$ and $W_2 \rightarrow W_1$ in Theorem~\ref{theorem:SINRcov}, and with some transformations in interference region.
\end{IEEEproof}

The investigation of two-tier Sub-6GHz network case aims to give a theoretical result in whether deploying mmWave SCells in hotspot regions could improve coverage performance by contrast with Sub-6GHz SCells. In Section~\ref{subsec:sinrCoverageProbability} we compare the coverage probability under several deployment scenarios, which shows that benefitting from the cancellation of inter-tier interference, the joint deployment of Sub-6GHz and mmWave BSs will achieve the best coverage performance in hotspot regions.

\subsection{Rate Analysis}\label{subsec:RateThroughputAnalysis}
Now we investigate the average achievable rate $\mathcal{R}\triangleq \mathbb{E}\left[ W\log_2\left( 1+\sinr \right)\right]$, which is defined as the Shannon bound for the SINR experienced over a cell, measured in bps. As the UEs are not uniformly distributed in the plane, $\mathcal{R}$ can be formulated by
\begin{align}\label{eq:rate_R_v0}
	\mathcal{R}
	= \mathbb{E}_{V_0} \left[ \mathcal{R}\left(v_0\right) \right]
	= \mathbb{E}_{V_0} \mathbb{E}\left[\left. W\log_2\left( 1+\sinr \right) \right|V_0=v_0 \right],
\end{align}
where $\mathcal{R}\left(v_0\right)$ is the average achievable rate for the UEs with distance $v_0$ to their cluster centers.

\begin{theorem}\label{theorem:aar}
The average achievable rate $\mathcal{R}$ is
\begin{align}\label{rate_expression}
	\mathcal{R}
	&= \int_0^\infty
	f_{V_0}\left(v_0\right) \left[
	\sum_{ k\in\left\{1,2\right\} }
	\mathcal{A}_k^\mathsf{c}\left( v_0 \right) W_k
	\int_0^\infty
	\mathcal{C}_k\left( 2^\rho-1;v_0 \right)
	\,\mathrm{d}\rho \right]
	\,\mathrm{d}v_0.
\end{align}
\end{theorem}

\begin{IEEEproof}
According to \eqref{eq:rate_R_v0}, the distance dependent rate $\mathcal{R}\left(v_0\right)$ is evaluated as
\begin{align}\label{eq:R_v0}
	\mathcal{R}\left(v_0\right)
	&= \mathbb{E}_\sinr\big[
	\left. W\log_2\left( 1+\sinr \right) \right|V_0=v_0
	\big]
	\nonumber
	\\
	&= \sum_{\mathclap{ k\in\left\{1,2\right\} }}
	\mathcal{A}_k^\mathsf{c}\left(v_0\right) W_k
	\mathbb{E}_{\sinr_k} \big[
	\left. \log_2\left( 1+\sinr_k \right) \right|V_0=v_0 \big]
	\nonumber
	\\
	&\overset{(a)}{=} \sum_{\mathclap{ k\in\left\{1,2\right\} }}
	\mathcal{A}_k^\mathsf{c}\left(v_0\right) W_k
	\int_0^\infty
	\mathbb{P}\left[ \sinr_k>2^\rho-1 \right]
	\,\mathrm{d}\rho,
\end{align}
where~(a) follows from $\mathbb{E}\left[X\right] = \int_0^\infty \mathbb{P}\left[X>x\right]\, \mathrm{d}x$ for a positive random variable $X$. The proof can be finished by taking expectation of $\mathcal{R}\left(v_0\right)$ with respect to $V_0$.
\end{IEEEproof}

\subsection{Logic Flow Diagram of Analysis}\label{subsec:logic_flow}
\begin{figure}[!t]
	\centering
	\includegraphics[width=4in]{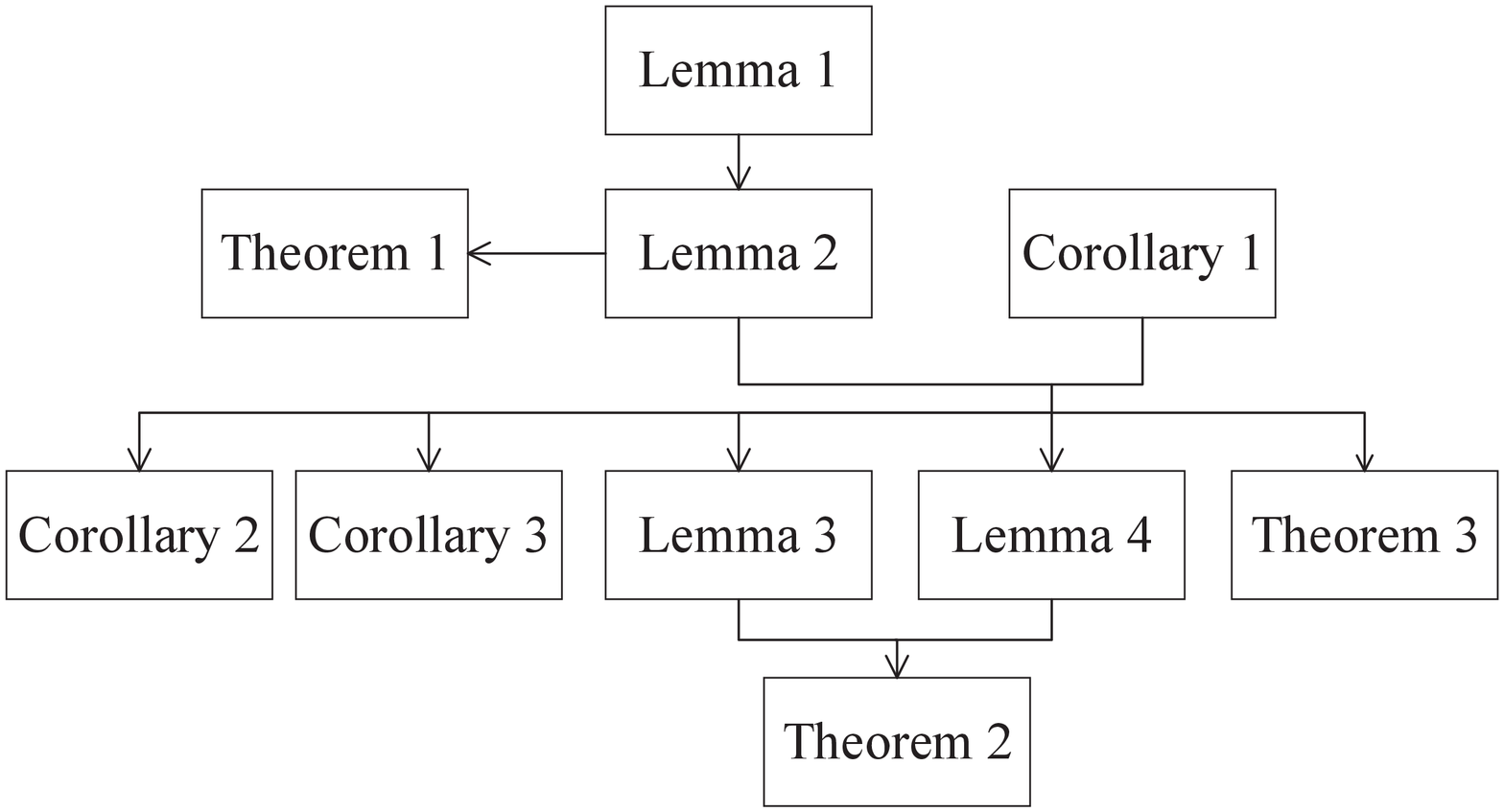}
	\caption{Logic flow diagram of analysis.}
	\label{flow}
\end{figure}

The relationship of lemmas, theorems and corollaries in this section is shown in Fig.~\ref{flow}. 
We first derive the distribution of nearest distance $R_k$ in Lemma~\ref{lemma:minDistancePDF}, based on which we get the expressions of conditional association probability $\mathcal{A}_k^c\left(v_0\right)$ in Lemma~\ref{lemma:A_c}. 
The distribution of conditional distance $X_k$ is derived in Corollary~\ref{corollary:Xk}, which is substantial in deriving the expressions of SINR coverage probabilities as well as average achievable rate in Lemma~\ref{lemma:C_1_con}, Lemma~\ref{lemma:C_2_con}, Theorem~\ref{theorem:SINRcov}, Theorem~\ref{theorem:aar}, Corollary~\ref{corollary:2:Case1} and Corollary~\ref{corollary:3:Case2}.

\section{Results and Discussions}\label{sec:simulation}
In this section, simulation and numerical results are presented to validate the accuracy of our theoretical analysis and provide useful insights into the network deployment of the integrated Sub-6GHz-mmWave cellular networks. For the numerical evaluation, we first model the network with one tier of sparsely deployed Sub-6GHz BSs and another tier of densely deployed mmWave BSs, where the Sub-6GHz BSs are distributed as a PPP and the mmWave BSs are distributed as a PCP. The locations of UEs are modeled as a PCP sharing same parent points with mmWave BSs. Sub-6GHz signals are assumed to experience single-slope path loss, Rayleigh fading and omnidirectional antenna gain, while mmWave signals follows blockage decided path loss and Nakagami fading, and directional beamforming gain. For the numerical results, we compute association probability and coverage probability by Monte Carlo simulation with $10^5$ iterations, where the BSs and UEs are generated in a circular shaped simulation area with radius $30$~km.
The detailed notations and values employed in the simulations are summarized in Table \ref{table:parameter}.

\begin{table}[!t]
	\centering\caption{Notations and Default Simulation Values}\label{table:parameter}
	\begin{tabular}{ccc}
		\hline
		\bfseries Notation & \bfseries Description & \bfseries Value\\
		\hline
		$\Phi_1,\Phi_p$						&	Sets of PPP deployed Sub-6GHz BSs and hotspot centers & \\
		\hline
		$\Phi_2,\Phi_u$						&	Sets of PCP deployed mmWave BSs and UEs & \\
		\hline
		$\lambda_1,\lambda_p$						&	Densities of Sub-6GHz BSs and hotspot centers	&	$30\,\text{/km}^2$, $5\,\text{/km}^2$\\
		\hline
		$n_\bs$								&	Number of mmWave BSs in each hotspot	&	$10$\\
		\hline
		$\sigma_\bs,\sigma_\ue$			&	Distribution standard deviations of mmWave BSs and UEs in traffic hotspots	&	$100$, $150$\\
		\hline
		$P_1,P_2$									&	Transmit power of Sub-6GHz BSs and mmWave BSs	&	$40\,\text{dBm}$, $30\,\text{dBm}$\\
		\hline
		$G_\mathsf{M},G_\mathsf{m},\theta_\mathsf{b}$			&	Parameters of sectored antenna model	&	$18\,\text{dBi}$, $-2\,\text{dBi}$, $10^\circ$\\
		\hline
		$p_\mathsf{L},R_\mathsf{B}$					&	Parameters of blockage ball model	&	$0.2$, $200~\text{m}$\\
		\hline
		$N_\mathsf{L},N_\mathsf{N}$					&	Nakagami-$m$ fading parameters for mmWave LoS and NLoS signals	&	$3$, $2$ \cite{Bai2014CM_mm}\\
		\hline
		$C_1,C_\mathsf{L},C_\mathsf{N}$				&	Path loss intercepts for different tiers	&	$-38.5\,\text{dB}$, $-61.4\,\text{dB}$, $-72\,\text{dB}$ \cite{JSAC2014Akdeniz}\\
		\hline
		$\alpha,\alpha_\mathsf{L},\alpha_\mathsf{N}$&	Path loss exponents for different tiers	&	$3$, $2$, $2.92$ \cite{JSAC2014Akdeniz}\\
		\hline
		$W_1,W_2$					&	Bandwidth of Sub-6GHz and mmWave carriers	&	$20\,\text{MHz}$, $1\,\text{GHz}$\\
		\hline
		$\sigma_k^2$		& Noise power for Sub-6GHz and mmWave & $-174\,\text{dBm/Hz} + 10\log_{10}\left(W_k\right) + 10\,\text{dB}$\\
		\hline
	\end{tabular}
\end{table}

\subsection{The Effects of Hotspot Parameters}\label{subsec:SimulationAssociation}
\begin{figure}[!t]
	\centering
	\includegraphics[width=4in]{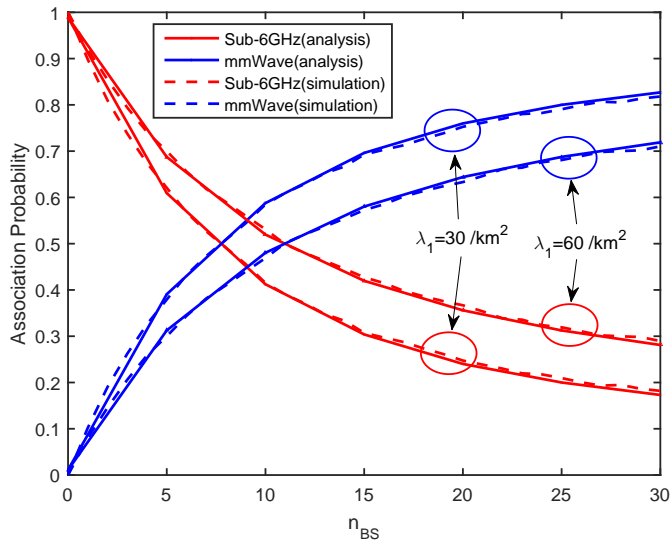}
	\caption{The association probability with variable number of mmWave BSs in each cluster for different values of $\lambda_1$~($\lambda_p=5~\text{/km}^2$, $\sigma_\bs=100$ and $\sigma_\ue=150$).}
	\label{fig2}
\end{figure}

The association probability with the variable clustered mmWave BSs $n_\bs$ is shown in Fig.~\ref{fig2}. It can be seen that the analytical and simulation results match well. The association probability of mmWave BSs monotonically increases with $n_\bs$, and this can be explained by the fact that the increase of $n_\bs$ leads to higher density of mmWave BSs and lower distance from the typical UE to the nearest mmWave BS, which improves the average received power of the typical UE from mmWave BSs. Note that mmWave BSs could achieve comparable association probability with Sub-6GHz BSs around $n_\bs=10$ in Fig. \ref{fig2}.

\begin{figure}[!t]
	\centering
	\includegraphics[width=4in]{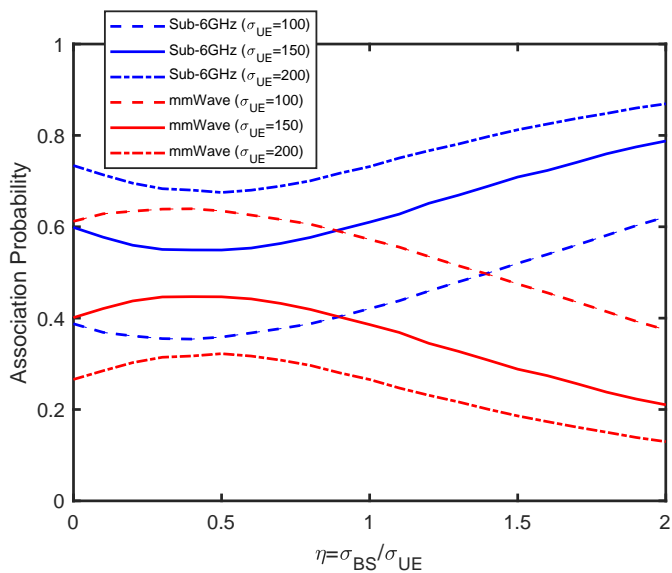}
	\caption{The association probability with variable distribution standard deviation ratio for different values of $\sigma_\ue$~($\lambda_1=30~\text{/km}^2$, $\lambda_p=5~\text{/km}^2$, $\sigma_\bs=100$ and $n_\bs=10$)}
	\label{fig3}
\end{figure}

Fig.~\ref{fig3} shows the effect of distribution standard deviation ratio $\eta$ on the association probability. The association probability of mmWave BSs increases slightly at first and then starts decreasing beyond $\eta=0.5$. Note that $\sigma_\bs=0$ corresponds to the case that the mmWave BSs are rightly located at the points of hotspot centers, and $\sigma_\bs=\infty$ corresponds to the case that the mmWave BSs are approximately independently distributed in $\mathbb{R}^2$. Since both the cases keep the mmWave BSs away from UEs, there exists an optimal value $\eta^\ast$ that maximizes the association probability of mmWave BSs. By adopting the proposed model and default values in Table~\ref{table:parameter}, it can be found that $\eta^\ast\approx 0.5$, which implies that the mmWave BSs in each cluster should be neither too aggregative nor too dispersal so that they can bring the largest performance enhancement.

\begin{figure}[!t]
	\centering
	\subfloat[5th and 50th percentile SINR]{\includegraphics[width=3in]{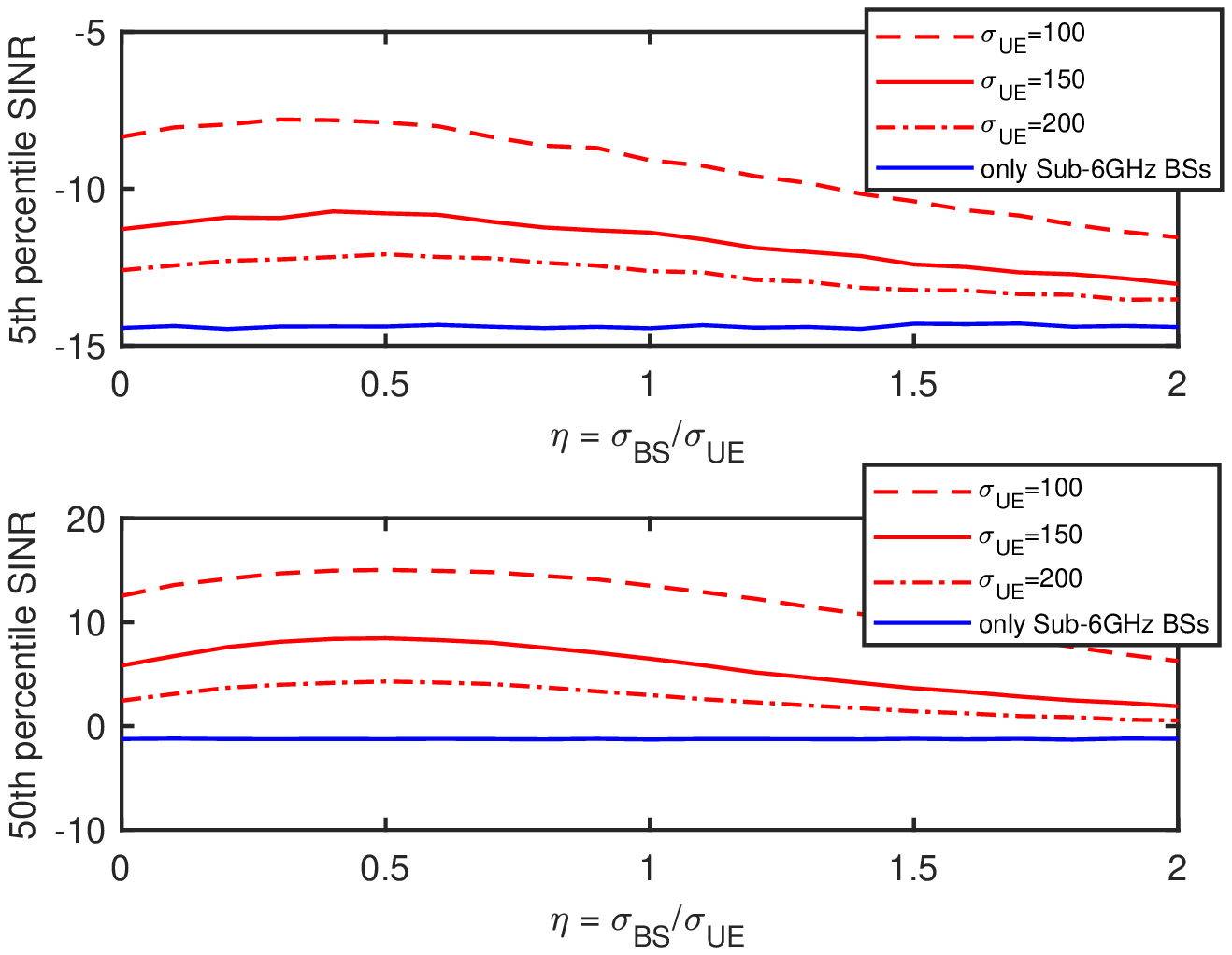}
	\label{fig4a}}
	\hfil
	\subfloat[5th and 50th percentile rate]{\includegraphics[width=3in]{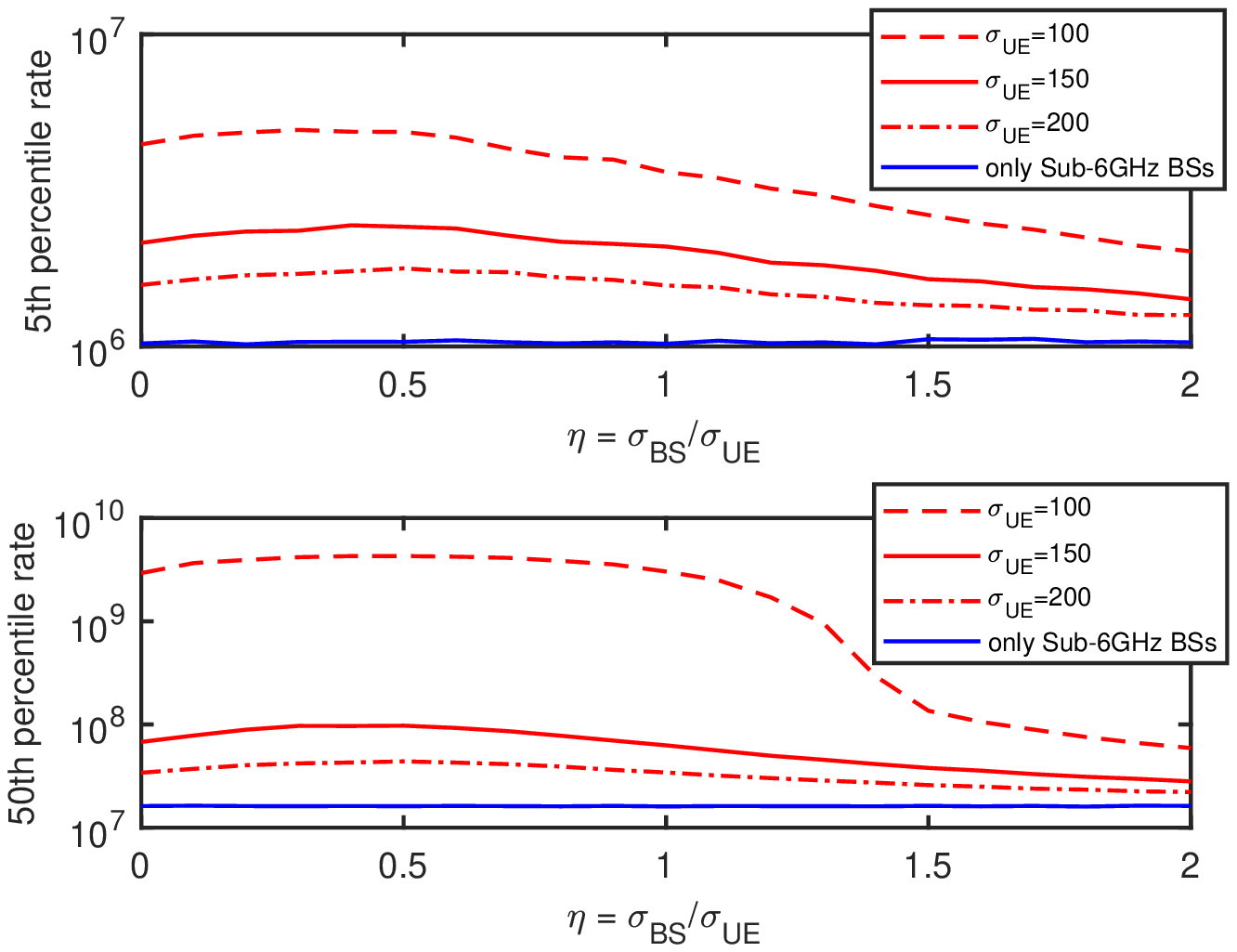}
	\label{fig4b}}
	\caption{The simulation results of cell edge and median performance with variable distribution standard deviation ratio for different values of $\sigma_\ue$~($\lambda_1=30~\text{/km}^2$, $\lambda_p=5~\text{/km}^2$, $\sigma_\bs=100$ and $n_\bs=10$). The result for single tier Sub-6GHz BS scheme is displayed as a baseline to show the benefit of mmWave BSs.}
	\label{fig4}
\end{figure}

To investigate the network performance against distribution standard deviation ratio $\eta$, we present the 5th/50th percentile SINR and 5th/50th percentile rate results in Figs.~\ref{fig4a} and \ref{fig4b}, respectively. The 5th percentile SINR~(edge SINR) and the 50th percentile SINR~(median SINR) are defined as $\left\{ \left.\tau\right|\mathcal{C}\left(\tau\right)=95\% \right\}$ and $\left\{ \left.\tau\right|\mathcal{C}\left(\tau\right)=50\% \right\}$, respectively, and so do the 5th/50th percentile rates. As a baseline, we provide aforementioned performance results with only PPP distributed Sub-6GHz BSs. The comparison of two schemes in Fig.~\ref{fig4} justify the performance gain brought by mmWave BSs. 
Additionally, the increase of $\sigma_\ue$, which corresponds to the expansion of UEs, will move the UEs away from hotspot centers and degrade the network performance. It is worth noting that all the four metrics increase firstly, peaking around $\eta^\ast\approx 0.5$, and then start decreasing. The optimal distribution standard deviation ratio $\eta^\ast$ that maximizes the network performance is around $0.5$, regardless of the variation of $\sigma_\ue$. This result indicates that we can optimize the deployment of mmWave BSs following Gaussian distribution with optimal $\sigma_\bs=\eta^\ast \sigma_\ue$.

\subsection{The Effects of Bias Values}\label{subsec:biaseffect}
\begin{figure*}[!t]
	\centering
	\subfloat[Association probability]{\includegraphics[width=3in]{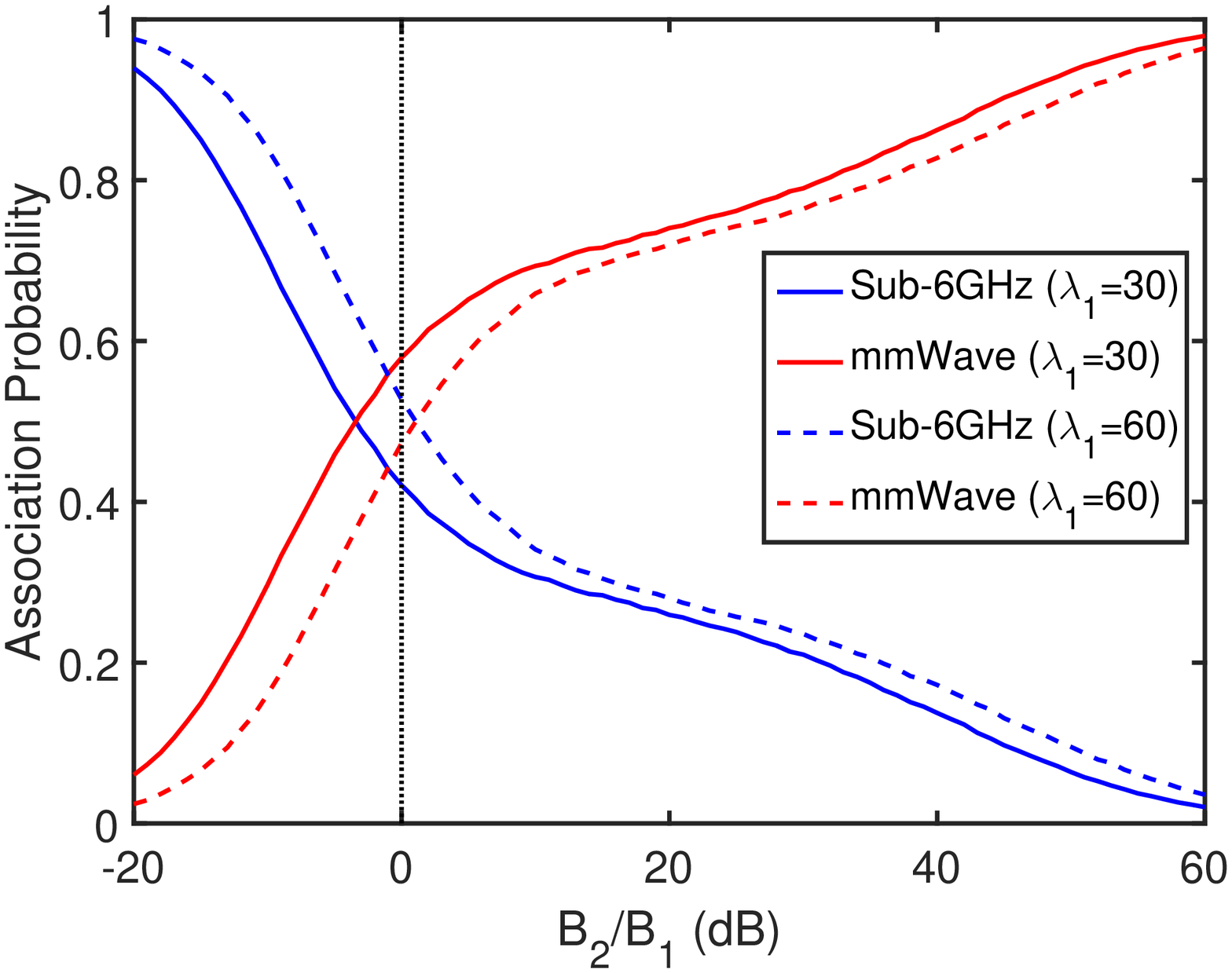}
	\label{fig5a}}
	\hfil
	\subfloat[5th and 50th percentile SINR]{\includegraphics[width=3in]{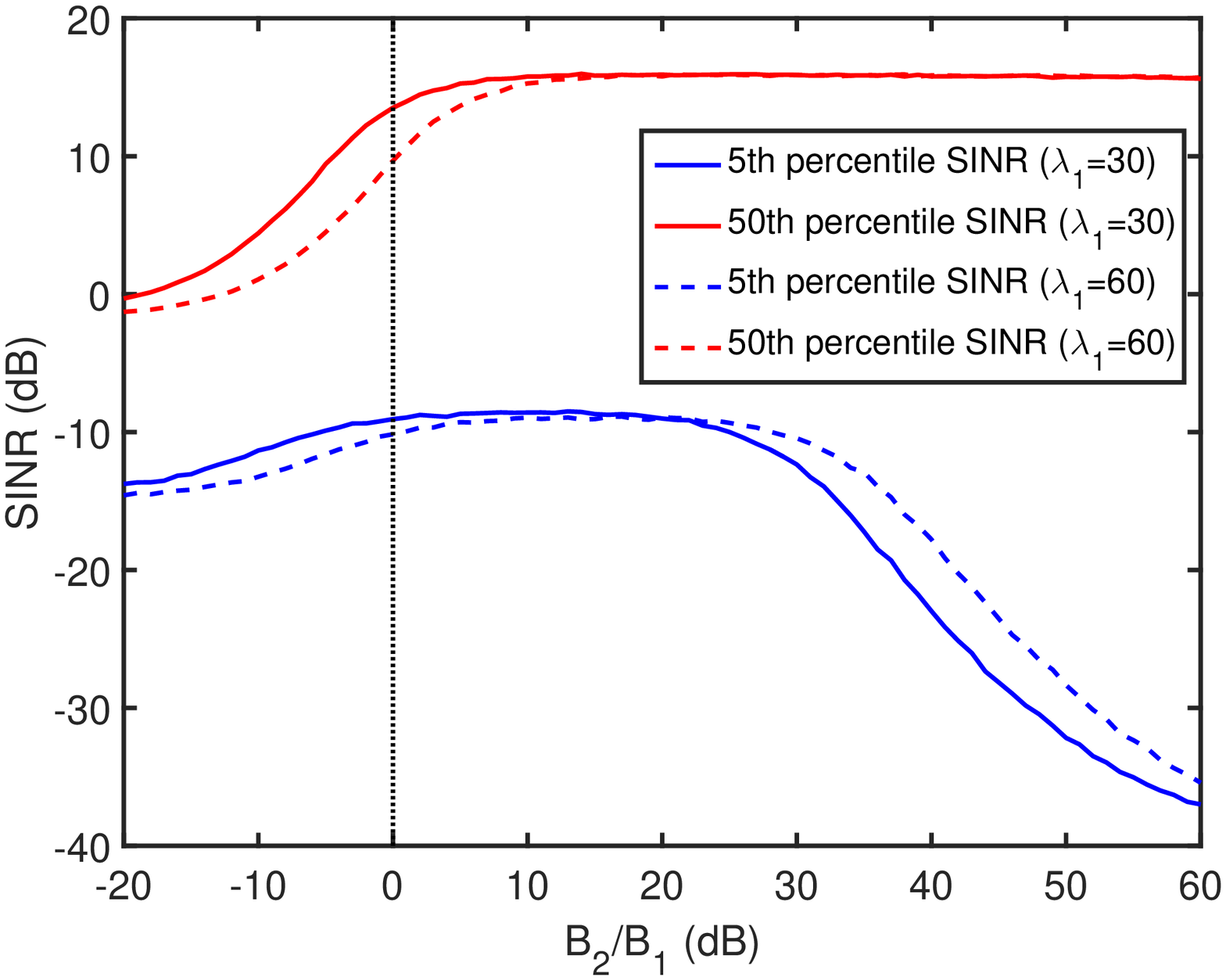}
	\label{fig5b}}
	\hfil
	\subfloat[5th and 50th percentile rate]{\includegraphics[width=3in]{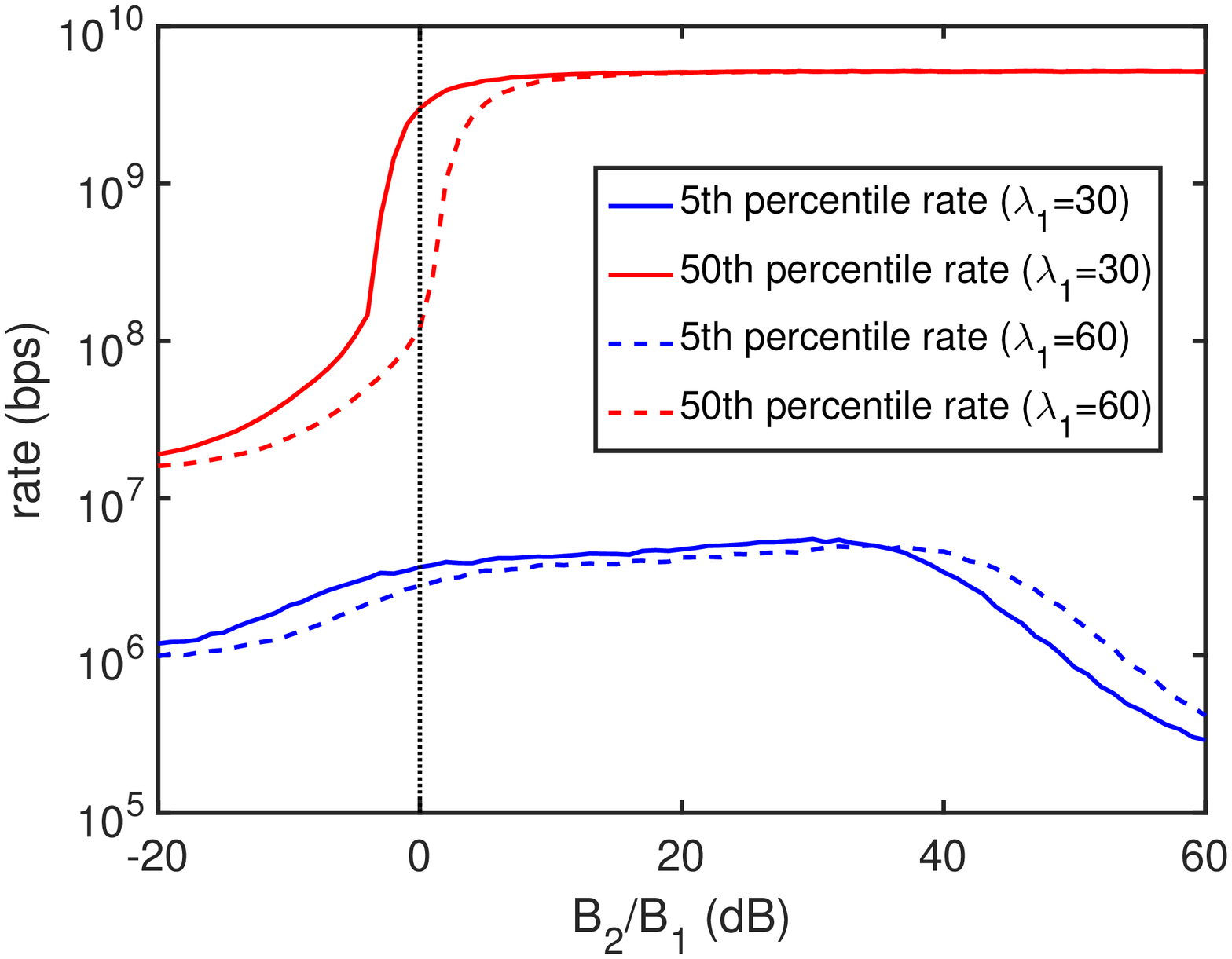}
	\label{fig5c}}
	\caption{The network performance with variable bias values~($\lambda_p=5~\text{/km}^2$, $n_\bs=10$, $\sigma_\bs=100$ and $\sigma_\ue=100$).}
	\label{fig5}
\end{figure*}

The bias value is an important parameter in adjusting the association probability of different network tiers and achieving the load balance. Since two different tiers are considered in the network model, we need only to investigate the network performance against bias ratio $B_2/B_1$, as shown in Fig.~\ref{fig5}. 
From Fig.~\ref{fig5a}, we observe that the increase of the bias ratio leads to higher association probability of mmWave BSs. Since the traffic hotspots are not fully covered in the plane, the association probability of mmWave SCells will achieve $90$\% under an extremely high bias ratio~(about $50$~dB). 
Moreover, it can be seen from Figs.~\ref{fig5b} and \ref{fig5c} that the median SINR and median rate monotonically increase in $\left[-20~\text{dB}, 10~\text{dB}\right]$ and saturate to a constant in $\left[20~\text{dB}, 60~\text{dB}\right]$, which is due to the excellent short-distance propagation and broad bandwidth at mmWave. When the bias ratio is below $10$~dB, the increase of bias ratio leads to higher percentage of mmWave UEs, and the network performance will be improved consequently. However, when the bias ratio is extremely high, some UEs that are supposed to be better served by Sub-6GHz BSs will be forced to associated with mmWave BSs, which would result in the degradation of the edge SINR and edge rate in $\left[30~\text{dB}, 60~\text{dB}\right]$.

\subsection{Distance Dependent Performance}
\begin{figure}[!t]
	\centering
	\subfloat[Average serving distance]{\includegraphics[width=3in]{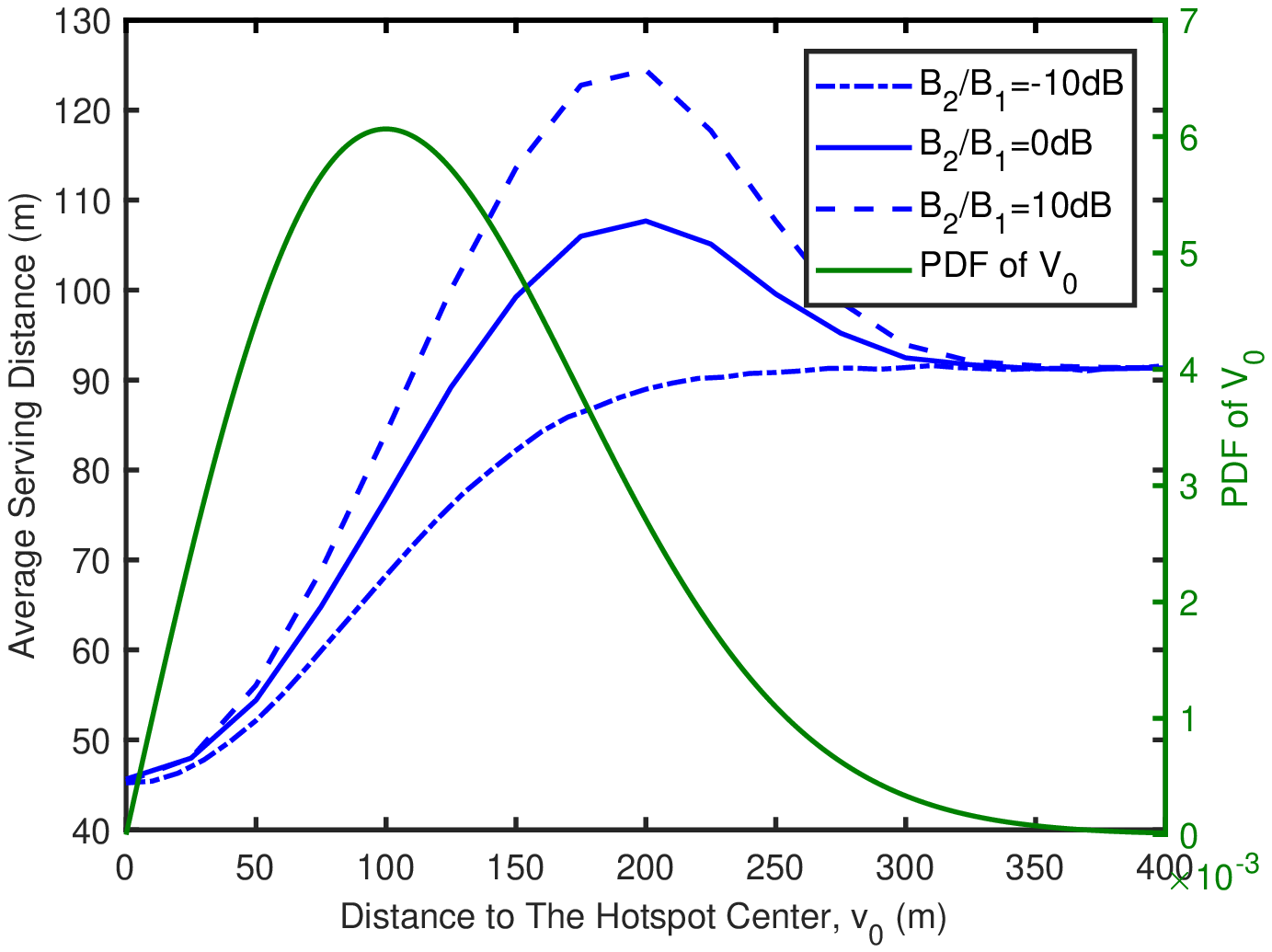}
	\label{fig6a}}
	\hfil
	\subfloat[Association probability]{\includegraphics[width=3in]{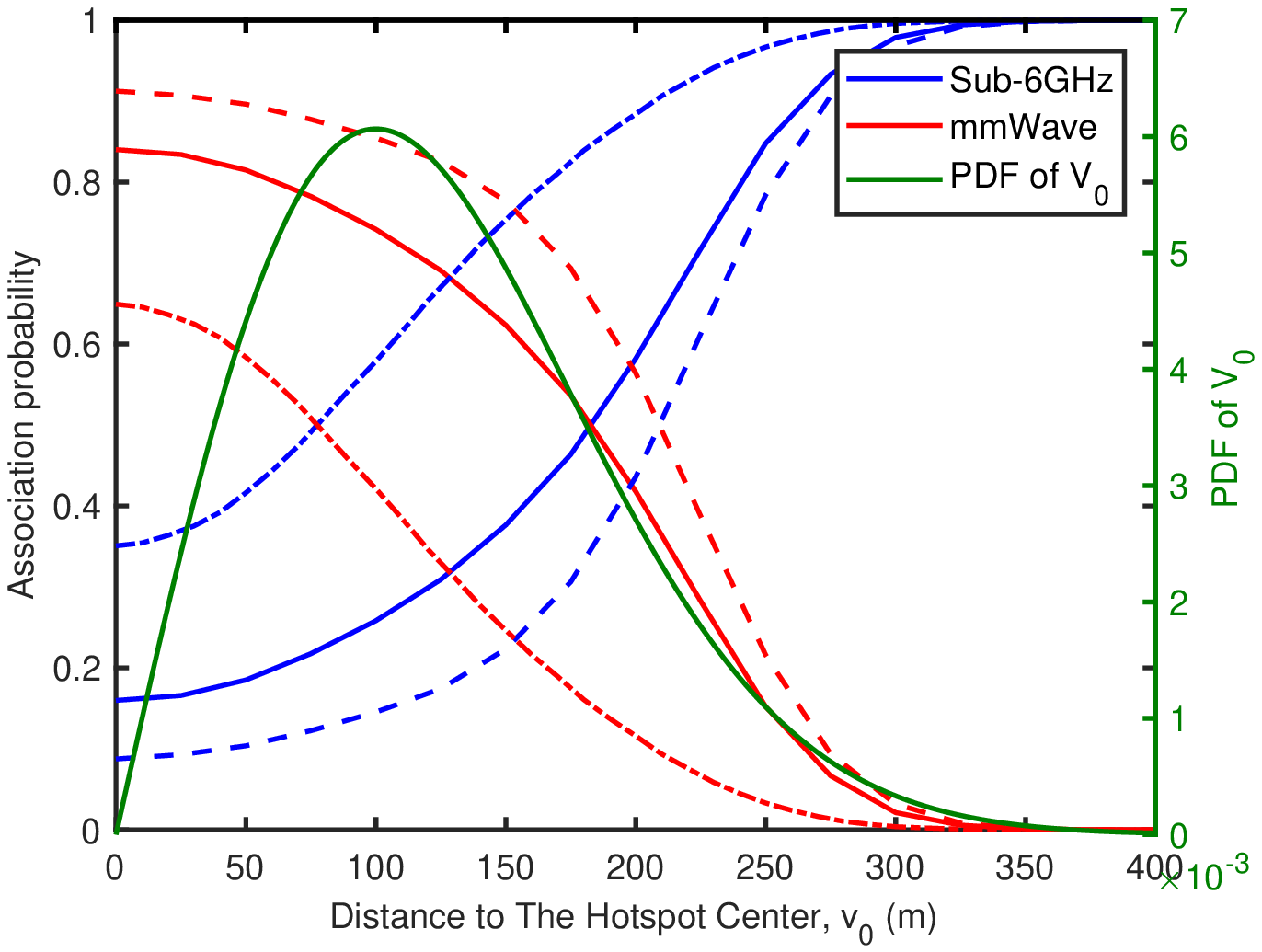}
	\label{fig6b}}
	\hfil
	\subfloat[5th and 50th percentile SINR]{\includegraphics[width=3in]{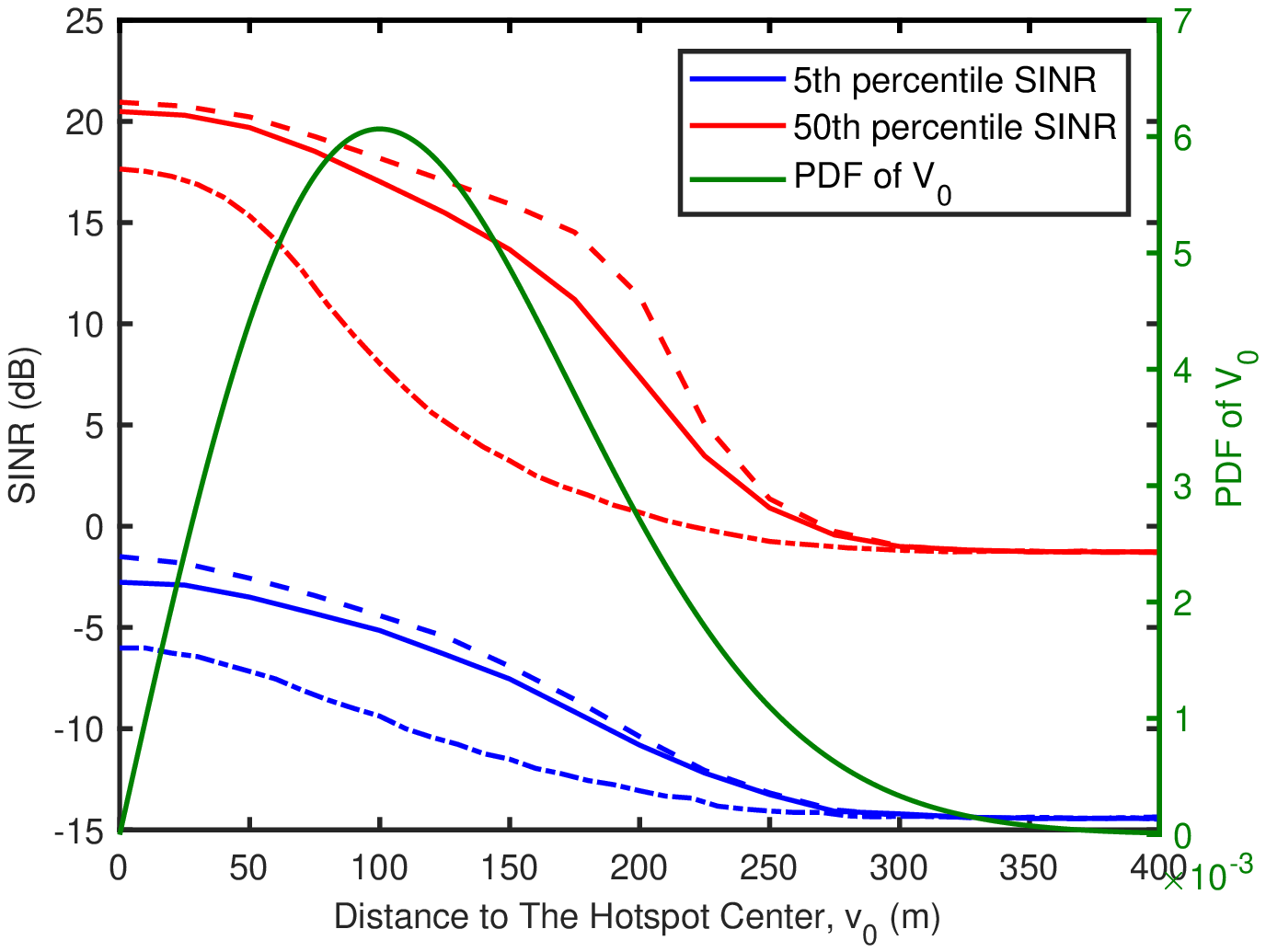}
	\label{fig6c}}
	\hfil
	\subfloat[5th and 50th percentile rate]{\includegraphics[width=3in]{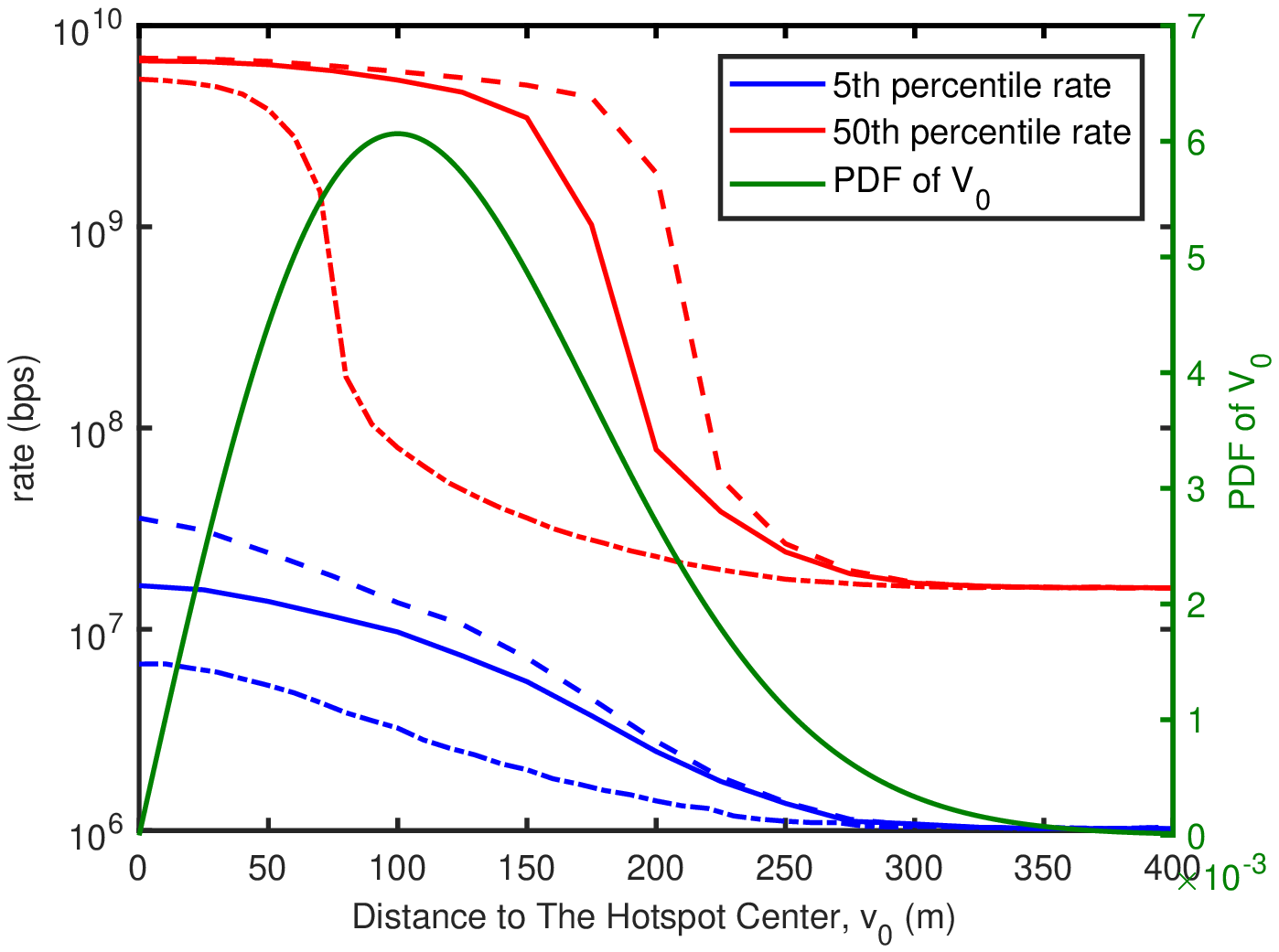}
	\label{fig6d}}
	\caption{The network performance with variable distance to the hotspot center for different bias ratios~($\lambda_1=30~\text{/km}^2$, $\lambda_c=5~\text{/km}^2$, $n_\bs=10$, $\sigma_\bs=50$ and $\sigma_\ue=100$).}
	\label{fig6}
\end{figure}

Since the adopted TCP model is not homogeneous in the plane, it is interesting to investigate the network performance for the UEs located at different regions, which is distinguished by the distance from UEs to their hotspot centers. We study four metrics including average serving distance, association probability, 5th/50th percentile SINR and 5th/50th percentile rate in Fig.~\ref{fig6}. As introduced in Section~\ref{sec:model}, UEs are distributed around the cluster centers following independent Gaussian distribution with standard deviation $\sigma_\ue$. It can be calculated that there are about $86.5$\% and $98.9$\% UEs located in $\cup_{\bm{c}\in\Phi_p} B(\bm{c},\sigma_\ue)$ and $\cup_{\bm{c}\in\Phi_p} B(\bm{c},2\sigma_\ue)$, respectively. The PDF of $V_0$ is also plotted in each subfigure of Fig.~\ref{fig6} for the ease of analysis.

From Fig.~\ref{fig6a}, it can be seen that as UEs are far away from hotspot centers, the average serving distances increase in $v_0\in\left[0, 200\right]$, and converge to a constant for long distances $v_0$. The reason is that the UEs near hotspot centers are more likely to be associated with mmWave BSs, as shown in Fig.~\ref{fig6b}, and UEs will deviate from the hotspot centers and gradually turn to be associated with Sub-6GHz MCells with the increase of $v_0$. Moreover, it can be seen from Fig.~\ref{fig6b} that with a higher bias ratio $B_2/B_1$, UEs will be more likely to be associated with mmWave BSs in clusters, which results in higher peak values of average serving distance in Fig.~\ref{fig6a}.

From Figs.~\ref{fig6c} and \ref{fig6d}, both the percentile SINR and rate monotonically decrease with $V_0$ and converge to constants, which reveals the fact that the UEs near hotspot centers could achieve better performance. For the UEs that far away from hotspot centers, they could only be associated with Sub-6GHz BSs, and the decline of SINR and rate will start later with a higher bias ratio.

\subsection{Coverage Probability}\label{subsec:sinrCoverageProbability}
\begin{figure}[!t]
	\centering
	\subfloat[single-tier scenario~(only mmWave BSs)]{\includegraphics[width=3in]{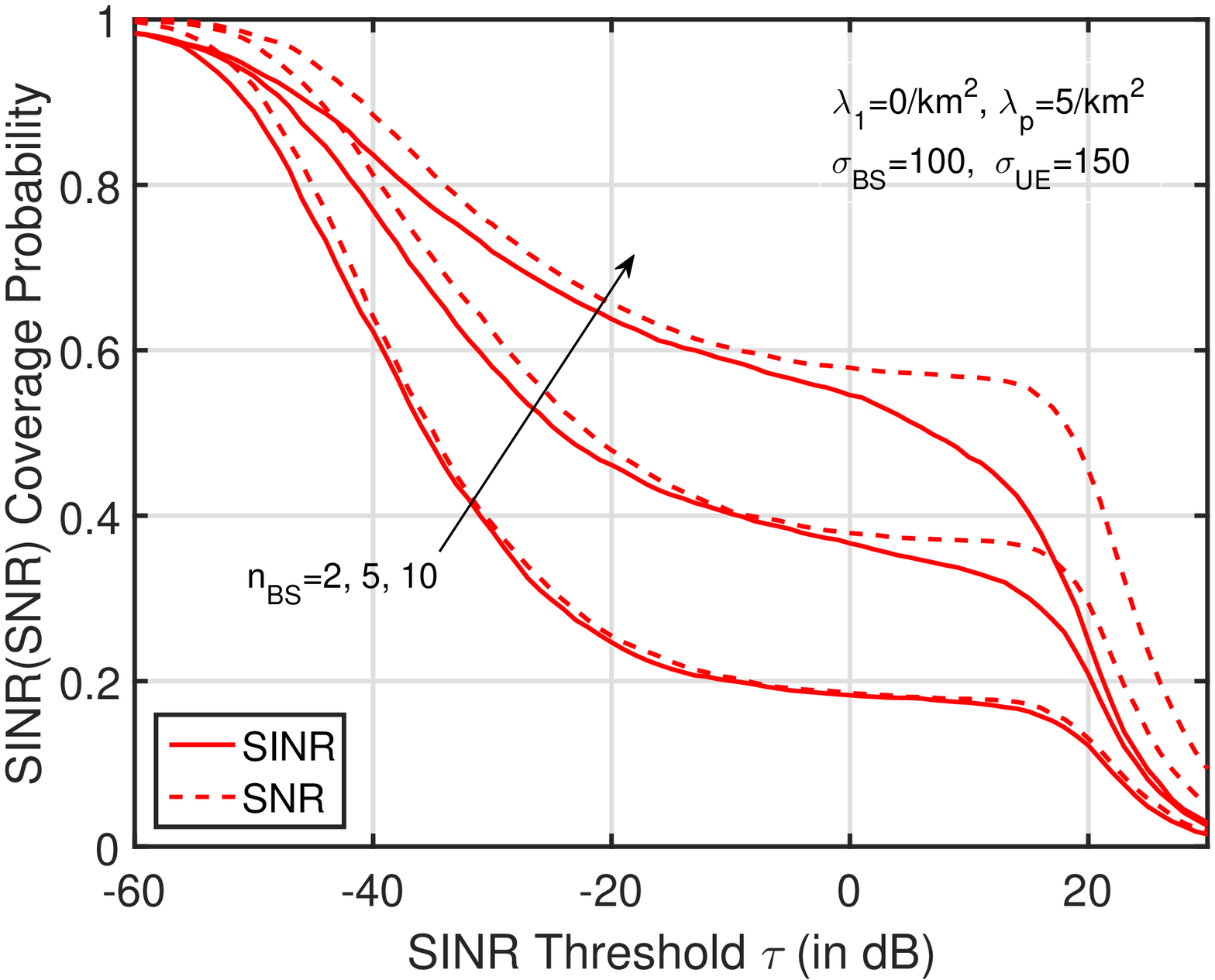}
	\label{fig7a}}
	\hfil
	\subfloat[two-tier scenario]{\includegraphics[width=3in]{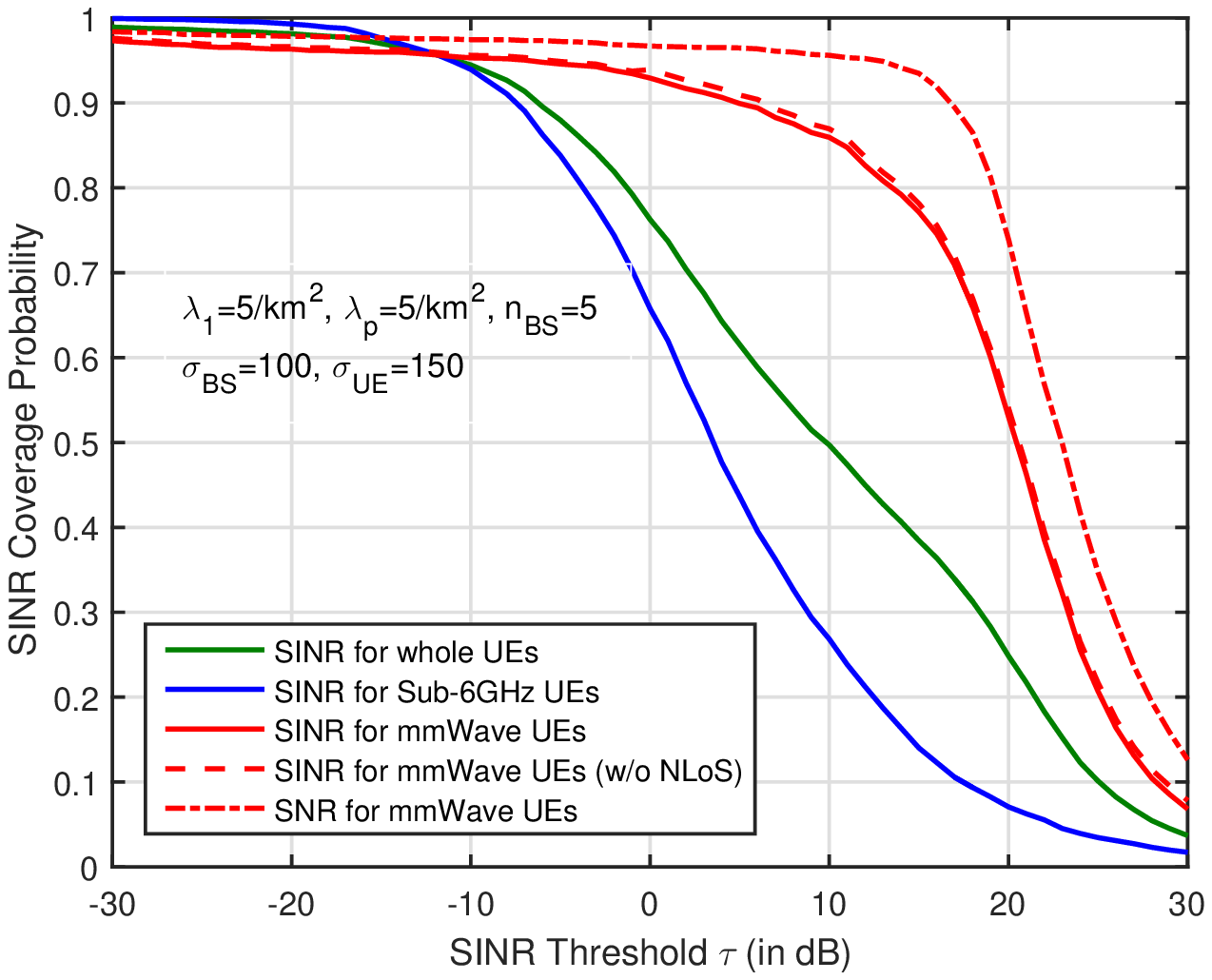}
	\label{fig7b}}
	\caption{The SINR coverage probability with variable threshold.}
	\label{fig7}
\end{figure}

The SINR coverage probability with different $n_\bs$ are shown in Fig.~\ref{fig7a} for a single-tier setting and in Fig.~\ref{fig7b} for a two-tier setting.
It can be seen from Fig.~\ref{fig7a} that the SINR and SNR are almost overlapping under $n_\bs=2$, whereas the gap between SINR and SNR becomes bigger with higher $n_\bs$. Similar results can be observed in Fig.~\ref{fig7b}. This can be explained by the fact that the network density in traffic hotspots is sufficiently high such that the mmWave interference has a significant impact on the coverage performance, which justifies the necessity of investigating SINR instead of SNR in the mmWave analysis.
The dashed line represents the approximate SINR coverage probability for mmWave UEs under the assumption that the interference of mmWave NLoS links is neglected, and it can be seen that it is close to the extract mmWave coverage result, which validates the tightness of the approximation in Corollary~\ref{corollary:2:Case1}. Moreover, as observed in Fig.~\ref{fig7b}, mmWave generally outperforms Sub-6GHz in SINR coverage performance, which is due to the centralized deployment and highly directional antennas of mmWave BSs.

\begin{figure}[!t]
	\centering
	\includegraphics[width=4in]{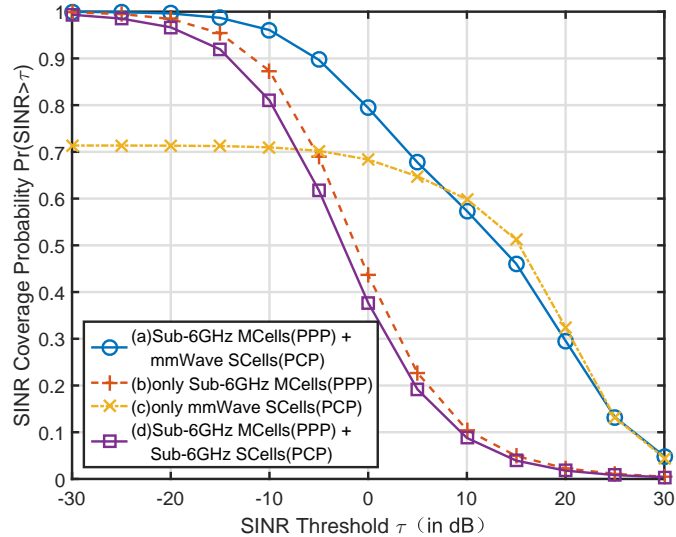}
	\caption{The SINR coverage probability with variable SINR threshold under four different deployments~(a), (b), (c) and (d).}
	\label{fig8}
\end{figure}

The SINR coverage probability $\mathcal{C}\left(\tau\right)$ with variable threshold $\tau$ under different deployments is shown in Fig.~\ref{fig8}. There are four settings: (a) the proposed two-tier network integrated with Sub-6GHz and mmWave; (b) with only Sub-6GHz BSs of (a); (c) with only mmWave BSs of (a); (d) the two-tier Sub-6GHz network introduced in Section~\ref{subsec:3_sp2}. In traffic hotspots, the joint deployment of Sub-6GHz and mmWave BSs, i.e., in setting~(a), can lead to the best coverage performance. Benefitting from the joint deployment of Sub-6GHz and mmWave BSs, there is no inter-tier interference in setting~(a), and thus the interference of setting~(a) is far less than that of setting~(d). In detail, we can see that under the target SINR threshold $0$~dB, the network with setting~(a) can achieve $\mathcal{C}^\text{(a)}\left(0~\text{dB}\right)\approx 80\%$ whereas the network with setting (d) can only achieve $\mathcal{C}^\text{(d)}\left(0~\text{dB}\right)\approx 40\%$. In setting~(c), with only PCP distributed mmWave BSs, due to the less of universal coverage provided by Sub-6GHz BSs, the SINR coverage probability is lower than that in setting~(a). Note that the SINR coverage probability of setting~(c) keeps at about $70$\% even at extremely low SINR threshold $\left[-30~\text{dB}, -10~\text{dB}\right]$, which reveals that there are some regions that far from hotspot centers are never covered. This also shows the importance of deploying Sub-6GHz together with mmWave.

\section{Conclusion}\label{sec:conclusion}
In this paper, we have proposed an analytical framework to analyze the performance of integrated Sub-6GHz-mmWave cellular network with traffic hotspots. The mmWave BSs are deployed in traffic hotspots to provide high data rate. We have derived the expressions of the association probability, SINR coverage probability and average achievable rate, and have investigated the network performance under different deployment schemes. The results reveal that deploying mmWave BSs in traffic hotspots can lead to better coverage performance than stand alone Sub-6GHz network. Moreover, the distribution standard deviation as well as bias value of mmWave BSs are the key factors in improving the utility of mmWave band. The optimal distribution standard deviation ratio is shown to be around $0.5$, and the bias of each tier need to be designed properly. The accuracy of our analysis has been validated through Monte Carlo simulations.

\appendices

\section{Proof of Lemma \ref{lemma:minDistancePDF}}\label{appendix:minDistancePDF}
For $k=1$, since $\Phi_1$ follows the homogeneous PPP with density $\lambda_1$, the CDF of $R_1$ can be evaluated as
\begin{align}\label{eq:appdA:CDF_R1}
	F_{R_1}\left(r;v_0\right)
	&= 1 - \mathbb{P}\left[ \text{There are no Sub-6GHz BSs in $O\left(0,r\right)$} \right] \nonumber
	\\
	&= 1 - \exp\left( -\pi\lambda_1 r^2 \right).
\end{align}
From \eqref{eq:appdA:CDF_R1}, it is easy to derive the PDF of $R_1$, as shown in \eqref{eq:f_R}.

For $k=2$, let $S_\mathsf{L}$ denote the distance from the typical UE to a randomly chosen mmWave LoS BS in cluster $\mathcal{X}_{\bm{c}_0}$, the CDF of $S_\mathsf{L}$ is given by
\begin{align}\label{eq:Ff_Sk_proof}
	F_{S_\mathsf{L}}\left(r;v_0\right)
	&= 1 - \mathbb{P}\left[ \text{There are no mmWave LoS BSs in $O\left(0,r\right)$} \right]
	\nonumber
	\\
	\quad
	&= 1 - \int_{\mathbb{R}^2\backslash O\left(0,r\right)}
	f_{\bm{X}_0}\left( \bm{x}-\mathbf{c}_0 \right)
	\cdot P_\mathsf{L}\left( \lVert\bm{x}\rVert \right)
	\,\mathrm{d}\bm{x}
	\nonumber
	\\
	\quad
	&= \int_{O\left(0,r\right)}
	f_{\bm{X}_0}\left( \bm{x}-\mathbf{c}_0 \right)
	\cdot P_\mathsf{L}\left( \lVert\bm{x}\rVert \right)
	\,\mathrm{d}\bm{x}
	\nonumber
	\\
	\quad
	&= \int_0^r \frac{t}{2\pi\sigma_\bs^2}
	\exp\left( -\frac{t^2+v_0^2}{2\sigma_\bs^2} \right)
	P_\mathsf{L}\left( t \right)
	J\left( \frac{v_0 t}{\sigma_\bs^2} \right)
	\,\mathrm{d}t,
\end{align}
where $J\left(x\right) = \int_{-\pi}^\pi \exp\left(x\cos\theta\right)\,\mathrm{d}\theta$. From \eqref{eq:Ff_Sk_proof}, we can obtain the PDF of $S_\mathsf{L}$, as shown in \eqref{eq:PDF:SL}.
For $R_2$, the equation $F_{R_2}\left(r;v_0\right) = 1 - \left[ 1-F_{S_\mathsf{L}}\left(r;v_0\right) \right]^{n_\bs}$ holds.
As such, $f_{R_k}\left(r;v_0\right)$ is derived by taking the derivative of $F_{R_k}\left(r;v_0\right)$ with respect to $r$.

\section{Proof of Lemma \ref{lemma:C_1_con}}\label{appendix:C_1_con}
When the typical UE is associated with a Sub-6GHz BS, i.e., $k=1$, the conditional SINR coverage probability can be expressed as
\begin{align}
    \mathcal{C}_1\left( \tau;v_0 \right)
    &= \mathbb{P}\left[ \left.
    \frac{P_1 G_1 C_1 h }{\sigma_1^2 + I_1} > \tau
    \right| K=1,V_0=v_0 \right] \\
    &= \mathbb{P}\left[ \left.
    h > \frac{\tau x^{\alpha_1} \left(\sigma_1^2 + I_1\right)}{P_1 G_1 C_1}
    \right| K=1,V_0=v_0 \right] \\
    &= \mathbb{E}_x \left[
    \exp\left( \frac{\tau \sigma_1^2 x^{\alpha_1}}{P_1 G_1 C_1} \right)
    \mathcal{L}_{I_1}\left( \frac{x^{\alpha_1} \tau}{P_1 G_1 C_1};v_0,x \right) \right] \\
	&= \int_0^\infty
	f_{X_1}\left( x; v_0 \right)
	\exp\left( -\frac{x^\alpha\tau\sigma_1^2}{b_1} \right)
	\LT{I_1}{\frac{x^\alpha\tau}{b_1};v_0,x}
	\,\mathrm{d}x.
\end{align}
The deviation of $\mathcal{L}_{I_1}\left( \cdot;v_0,x \right)$ follows on the same lines as in \cite{andrews2011tractable}.

\section{Proof of Lemma \ref{lemma:C_2_con}}\label{appendix:C_L_con}
When the typical UE is associated with an mmWave LoS BS, i.e., $k=2$, the conditional SINR coverage probability can be expressed as
\begin{align}
	\mathcal{C}_2\left( \tau;v_0 \right)
	&= \mathbb{E}_{x,I_\mathsf{L}}
	\mathbb{P}\left[ \left.
	\frac{ P_2 G_\mathsf{M} h C_\mathsf{L} x^{-\alpha_\mathsf{L}} }{ I_\mathsf{L} + \sigma_\mathsf{mm}^2 } > \tau
	\right|V_0=v_0\right]
	\\
	&= \mathbb{E}_{x,I_\mathsf{L}}
	\mathbb{P}\left[ h >
	\underbrace{ \frac{x^{\alpha_\mathsf{L}}\tau\left(I_\mathsf{L} + \sigma_2^2\right)}{P_2 G_\mathsf{M} C_\mathsf{L}} }_{ J_\mathsf{L} }
	\Big|V_0=v_0\right].
\end{align}
According to the Alzer's inequality \cite{alzer1997some}, for a normalized gamma random variable $h$ with parameter $N_\mathsf{L}$, the probability $\mathbb{P}\left[h >J_\mathsf{L}\right]$ can be approximated as
\begin{align}
	\mathbb{P}\left[ h>J_\mathsf{L} \right]
	&\approx 1-\left( 1-e^{-\chi_2 J_\mathsf{L}} \right)^{N_\mathsf{L}}
	\nonumber
	\\
	&= \sum_{n=1}^{N_\mathsf{L}}
	\left(-1\right)^{n+1}
	\binom{N_\mathsf{L}}{n}
	e^{-\chi_2 n J_\mathsf{L}},
\end{align}
where $\chi_2=N_\mathsf{L} \left(N_\mathsf{L}!\right)^{-\frac{1}{N_\mathsf{L}}}$.
Thus, $\mathcal{C}_2\left(\tau;v_0\right)$ can be written as
\begin{align}\label{eq:CL_con_pre}
	\mathcal{C}_2\left( \tau;v_0 \right)
	&\approx \int_0^\infty
	f_{X_\mathsf{L}}\left( x; v_0\right)
	\sum_{n=1}^{N_\mathsf{L}}
	\left(-1\right)^{n+1} \binom{N_\mathsf{L}}{n}
	\mathbb{E}_{I_\mathsf{L}}
	\left[ e^{-\chi_2 n J_\mathsf{L}}\right]
	\,\mathrm{d}x
	\nonumber
	\\
	&= \int_0^\infty
	f_{X_\mathsf{L}}\left( x; v_0\right)
	\sum_{n=1}^{N_\mathsf{L}} a_2\left(n\right)
	\exp\left(
	-\frac{ \sigma_2^2 \chi_2 n x^{\alpha_\mathsf{L}} \tau }{b_2} \right)
	\LT{I_2}{
	\frac{ \chi_2 n x^{\alpha_\mathsf{L}} \tau }{b_2};v_0,x }
	\,\mathrm{d}x,
\end{align}
where $a_2\left(n\right) = \left(-1\right)^{n+1} \binom{N_\mathsf{L}}{n}$, and $b_2 = P_2 G_\mathsf{M} C_\mathsf{L}$.

Now, we calculate the Laplace transform of the interference $I_2$. As discussed in Section~\ref{subsec:SINRCoverageAnalysis}, $I_2$ can be expressed as $I_2=I_2^\mathsf{intra}+I_2^\mathsf{inter}$, and we have $\LT{I_2}{s;v_0,x}=\LT{I_2^\mathsf{intra}}{s;v_0,x} \cdot\LT{I_2^\mathsf{inter}}{s;v_0,x}$. The first term is computed as follows:
\begin{align}\label{eq:I_2_intra_L}
	&\mathcal{L}_{I_2^\intra}^\mathsf{L} \left({s;v_0,x}\right)
	\nonumber\\
	&= \mathbb{E}_{ \mathcal{X}_{\bm{c}_0},G_\mathsf{b},h } \left[
	\exp\left( -s \sum_{ \bm{x}\in\mathcal{X}_{\bm{c}_0}^\mathsf{L}\!\backslash\bm{x}_0^* }
	P_2 G_\mathsf{b} h_{\bm{x}}
	C_\mathsf{L} \lVert\bm{x}\rVert^{-\alpha_\mathsf{L}}
	\right) \right]
	\nonumber
	\\
	&= \mathbb{E}_{ \mathcal{X}_{\bm{c}_0} } \left\{
	\prod_{\scriptstyle{ \bm{x}\in\mathcal{X}_{\bm{c}_0}^\mathsf{L}\!\backslash\bm{x}_0^* } }
	\mathbb{E}_{G_\mathsf{b},h} \Big[
	\exp\left( -s P_2 G_\mathsf{b} h_{\bm{x}}
	C_\mathsf{L} \lVert\bm{x}\rVert^{-\alpha_\mathsf{L}} \right)
	\Big] \right\}
	\nonumber
	\\
	&\overset{(a)}{=} \mathbb{E}_{\mathcal{X}_{\bm{c}_0}} \left\{
	\prod_{\scriptstyle{ \bm{x}\in\mathcal{X}_{\bm{c}_0}^\mathsf{L}\!\backslash\bm{x}_0^* } }
	\mathbb{E}_{G_\mathsf{b}} \left[
	\left( 1 + s P_2 G_\mathsf{b} C_\mathsf{L} \lVert\bm{x}\rVert^{-\alpha_\mathsf{L}} \right)^{-N_\mathsf{L}}
	\right] \right\}
	\nonumber
	\\
	&\overset{(b)}{=} \exp\left\{ -2\pi\left(n_\bs-1\right)
	\int_x^\infty
	f_{X_2}\left(r;v_0\right)
	\mathbb{E}_{G_\mathsf{b}}
	\left[ 1 - \left(
	1 + s P_2 G_\mathsf{b} C_\mathsf{L} r^{-\alpha_\mathsf{L}} \right)^{-N_\mathsf{L}}
	\right] r
	\,\mathrm{d}r \right\},
\end{align}
where (a) follows from the moment generating function of normalized Gamma variable $h$, and (b) follows from the probability generating functional (PGFL) of Poisson process $\mathcal{X}_{\bm{c}_0}^\mathsf{L}\!\backslash\bm{x}_0^*$ with intensity measure $\left(n_\bs-1\right) f_{X_2}\left(r;v_0\right)$. Following the same procedures as \eqref{eq:I_2_intra_L}, we have
\begin{align}\label{eq:I_2_intra_N}
	&\mathcal{L}_{I_2^\intra}^\mathsf{N} \left({s;v_0,x}\right)
	= \exp\left\{ -2\pi\left(n_\bs-1\right)
	\int_{ \delta_{2,\mathsf{N}}\left(x\right) }^\infty
	\!\!f_{X_2}\left(r;v_0\right)
	\right.\nonumber
	\\
	&\qquad\qquad\qquad\qquad \left.\times
	\mathbb{E}_{G_\mathsf{b}}
	\left[ 1 - \left(
	1 + s P_2 G_\mathsf{b} C_\mathsf{N} r^{-\alpha_\mathsf{N}} \right)^{-N_\mathsf{N}}
	\right] r
	\,\mathrm{d}r \right\}.
\end{align}
Hence, $\LT{I_2^\mathsf{intra}}{s;v_0,x}$ can be expressed as
\begin{align}\label{eq:I_2_intra}
	\LT{I_2^\intra}{s;v_0,x}
	&= \mathcal{L}_{I_2^\intra}^\mathsf{L} \left({s;v_0,x}\right)
	\cdot\mathcal{L}_{I_2^\intra}^\mathsf{N} \left({s;v_0,x}\right)
	\nonumber
	\\
	&= \exp\Bigg\{ -2\pi\left(n_\bs-1\right)
	\sum_{ i\in\left\{\mathsf{L,N}\right\} }
	\int_{ \delta_{\mathsf{L},i}\left(x\right) }^\infty
	f_{X_2}\left(r;v_0\right)
	\nonumber
	\\
	&\qquad \times\left[
	1 - \sum_{ \mathclap{j\in\left\{\mathsf{m,M}\right\}} }
	p_j \left( 1 + s P_2 G_j C_i r^{-\alpha_\mathsf{L}} \right)^{-N_\mathsf{L}}
	\right] r
	\,\mathrm{d}r \Bigg\}.
\end{align}

Furthermore, the second term can be derived by leveraging the result of $\LT{I_2^\mathsf{intra}}{s;v_0,x}$, and
\begin{align}\label{eq:I_2_inter}
	\LT{I_\mathsf{L}^\inter}{s;v_0,x}
	&= \mathbb{E}_{\Phi_p} \left[
	\prod_{\bm{c}\in\Phi_p}
	\LT{\left.I_2^\mathsf{intra}\right|\bm{c},n_\bs+1}{s;\lVert\bm{c}\rVert,0}
	\right]
	\nonumber\\
	&\overset{(a)}{=} \exp\left\{ -2\pi\lambda_p
	\int_0^\infty
	\left[ 1 - \LT{\left.I_\mathsf{L}^\intra\right|n_\bs+1}{s;v,0} \right] v
	\,\mathrm{d}v \right\},
\end{align}
where (a) follows from the PGFL of homogeneous PPP $\Phi_p$. The proof is finished by substituting \eqref{eq:I_2_intra} and \eqref{eq:I_2_inter} into \eqref{eq:CL_con_pre}.

\bibliographystyle{IEEEtran}

\begin{thebibliography}{99}
\bibitem{Access2013Rappaport}
T.~S.~Rappaport, S.~Sun, R.~Mayzus, H.~Zhao, Y.~Azar, K.~Wang, G.~N.~Wong, J.~K.~Schulz, M.~Samimi, and F.~Gutierrez, ``Millimeter wave mobile communications for 5G cellular: It will work!'' \emph{IEEE Access}, vol.~1, pp. 335--349, 2013.
\bibitem{rappaport2014millimeter}
T.~S.~Rappaport, R.~W.~Heath~Jr, R.~C.~Daniels, and J.~N.~Murdock, \emph{Millimeter wave wireless communications}. Pearson Education, 2014.
\bibitem{TWC2019Gao}
X.~Gao, P.~Wang, D.~Niyato, K.~Yang, and J.~An, ``Auction-based time scheduling for backscatter-aided RF-powered cognitive radio networks,'' in \emph{IEEE Trans. Wireless Commun.}, vol.~18, no.~3, pp. 1684--1697, Mar. 2019.
\bibitem{JSAC2017Yu_antenna}
X.~Yu, J.~Zhang, M.~Haenggi, and K.~B.~Letaief, ``Coverage analysis for millimeter wave networks: The impact of directional antenna arrays,'' \emph{IEEE J. Sel. Areas Commun.}, vol.~35, no.~7, pp. 1498--1512, Jul. 2017.
\bibitem{CM2018yang}
K.~Yang, N.~Yang, N.~Ye, M.~Jia, Z.~Gao, and R.~Fan, ``Non-orthogonal multiple access: achieving sustainable future radio access,'' \emph{IEEE Commun. Mag.}, vol.~57, no.~2, pp. 116--121, Feb. 2019.
\bibitem{TVT2019Hang}
H. Yuan, J. An, N. Yang, K. Yang, and T. Q. Duong, ``Low complexity hybrid precoding for multiuser millimeter wave systems over frequency selective channels,'' \emph{IEEE Trans. Veh. Technol.}, vol. 68, no. 1, pp. 983–987, Jan. 2019. 
\bibitem{TC2017mmTut_Andrews}
J.~G.~Andrews, T.~Bai, M.~N.~Kulkarni, A.~Alkhateeb, A.~K. Gupta, and R.~W. Heath, ``Modeling and analyzing millimeter wave cellular systems,'' \emph{IEEE Trans. Commun.}, vol.~65, no.~1, pp. 403--430, Jan. 2017.
\bibitem{glocom2016mminitial}
Y.~Li, J.~G.~Andrews, F.~Baccelli, T.~D.~Novlan, and J.~Zhang, ``On the initial access design in millimeter wave cellular networks,'' in \emph{Proc. IEEE Globecom Workshops (GC Wkshps)}, pp. 1--6, Dec. 2016.
\bibitem{twc2017initial}
A.~Alkhateeb, Y.~H. Nam, M.~S.~Rahman, J.~Zhang, and R.~W.~Heath, ``Initial beam association in millimeter wave cellular systems: Analysis and design insights,'' \emph{IEEE Trans. Wireless Commun.}, vol.~16, no.~5, pp. 2807--2821, May. 2017.
\bibitem{CST2017ElSawy_SGtut}
H.~ElSawy, A.~Sultan-Salem, M.~S.~Alouini, and M.~Z.~Win, ``Modeling and analysis of cellular networks using stochastic geometry: A tutorial,'' \emph{IEEE Commun. Surveys Tuts.}, vol.~19, no.~1, pp. 167--203, First quarter 2017.
\bibitem{andrews2011tractable}
J.~G.~Andrews, F.~Baccelli, and R.~K.~Ganti, ``A tractable approach to coverage and rate in cellular networks,'' \emph{IEEE Trans. Commun.}, vol.~59, no.~11, pp. 3122--3134, Nov. 2011.
\bibitem{kong2016jsacginibre}
H.~Kong, I.~Flint, P.~Wang, D.~Niyato, and N.~Privault, ``Exact performance analysis of ambient RF energy harvesting wireless sensor networks with Ginibre point process,'' \emph{IEEE J. Sel. Areas Commun.}, vol.~34, no.~12, pp. 3769--3784, Dec. 2016.
\bibitem{flint2017twcphchp}
I.~Flint, H.~Kong, N.~Privault, P.~Wang, and D.~Niyato, ``Analysis of heterogeneous wireless networks using Poisson hard-core hole process,'' \emph{IEEE Trans. Wireless Commun.}, vol.~16, no.~11, pp. 7152--7167, Nov. 2017.
\bibitem{TWC2016De_Elshaer}
H.~Elshaer, M.~N.~Kulkarni, F.~Boccardi, J.~G.~Andrews, and M.~Dohler, ``Downlink and uplink cell association with traditional macrocells and millimeter wave small cells,'' \emph{IEEE Trans. Wireless Commun.}, vol.~15, no.~9, pp. 6244--6258, Sep. 2016.
\bibitem{tcom2018yi}
W.~Yi, Y.~Liu, and A.~Nallanathan, ``Cache-enabled HetNets with millimeter wave small cells,'' \emph{IEEE Trans. Commun.}, vol.~66, no.~11, pp. 5497--5511, 2018.
\bibitem{icc2016omar}
M.~S.~Omar, M.~A.~Anjum, S.~A.~Hassan, H.~Pervaiz, and Q.~Niv, ``Performance analysis of hybrid 5G cellular networks exploiting mmwave capabilities in suburban areas,'' in \emph{Proc. ICC}, pp. 1--6, May. 2016.
\bibitem{vtc2016multibandWang}
F.~Wang, H.~Wang, H.~Feng, and X.~Xu, ``A hybrid communication model of millimeter wave and microwave in D2D network,'' in \emph{Proc. IEEE 83rd Veh. Technol. Conf. (VTC Spring)}, pp. 1--5, May. 2016.
\bibitem{saha2018pcp}
C. Saha, M. Afshang, and H. S. Dhillon, ``3GPP-inspired HetNet model using Poisson cluster process: sum-product functionals and downlink coverage,'' \emph{IEEE Trans. Commun.}, vol.~66, no.~5, pp. 2219--2234, 2018.
\bibitem{saha2018pcp0}
C. Saha, H. S. Dhillon, N. Miyoshi and J. G. Andrews, ``Unified Analysis of HetNets Using Poisson Cluster Processes Under Max-Power Association,'' \emph{IEEE Trans. Wireless Commun.}, vol.~18, no.~8, pp. 3797-3812, Aug. 2019.
\bibitem{ganti2009pcp}
R. K. Ganti and M. Haenggi, ``Interference and outage in clustered wireless Ad Hoc networks,'' \emph{IEEE Trans. Inf. Theory}, vol.~55, no.~9, pp. 4067--4086, 2009.
\bibitem{afshang2018}
M. Afshang, C. Saha, and H. S. Dhillon, ``Equi-Coverage Contours in Cellular Networks,'' \emph{IEEE Wireless Commun. Lett.}, vol. 7, no. 5, pp. 700--703, 2018.
\bibitem{ppp2016mankar}
P.~D.~Mankar, G.~Das, and S.~S.~Pathak, ``Modeling and coverage analysis of BS-centric clustered users in a random wireless network,'' \emph{IEEE Wireless Commun. Lett.}, vol.~5, no.~2, pp. 208--211, Apr. 2016.
\bibitem{kppppcp2016saha}
C.~Saha and H.~S.~Dhillon, ``Downlink coverage probability of K-tier HetNets with general non-uniform user distributions,'' in \emph{Proc. ICC}, pp. 1--6, May. 2016.
\bibitem{pcp20172tier}
M.~Afshang and H.~S.~Dhillon, ``A new clustered HetNet model to accurately characterize user-centric small cell deployments,'' in \emph{Proc. IEEE Wireless Commun. Netw. Conf. (WCNC)}, pp. 1--6, Mar. 2017.
\bibitem{glocom2015d2dpcpAfshang}
M.~Afshang, H.~S.~Dhillon, and P.~H.~J.~Chong, ``Coverage and area spectral efficiency of clustered device-to-device networks,'' in \emph{Proc. IEEE Globecom}, pp. 1--6, Dec. 2015.
\bibitem{afshang2016nncd}
M. Afshang, C. Saha, and H. S. Dhillon, ``Nearest-neighbor and contact distance distributions for Thomas cluster process,'' \emph{IEEE Wireless Commun. Lett.}, pp. 1--1, 2016.
\bibitem{miyoshi2019pcp}
N. Miyoshi, ``Downlink coverage probability in cellular networks with Poisson-Poisson cluster deployed base stations,'' \emph{IEEE Wireless Commun. Lett.}, vol.~8, no.~1, pp. 5--8, 2019.
\bibitem{Bai2014CM_mm}
T.~Bai, A.~Alkhateeb, and R.~W.~Heath, ``Coverage and capacity of millimeter-wave cellular networks,'' \emph{IEEE Commun. Mag.}, vol.~52, no.~9, pp. 70--77, Sep. 2014.
\bibitem{JSAC2015Singh_backhaul}
S.~Singh, M.~N.~Kulkarni, A.~Ghosh, and J.~G.~Andrews, ``Tractable model for rate in self-backhauled millimeter wave cellular networks,'' \emph{IEEE J. Sel. Areas Commun.}, vol.~33, no.~10, pp. 2196--2211, Oct. 2015.
\bibitem{TWC2012HetNet_Jo}
H.~S.~Jo, Y.~J.~Sang, P.~Xia, and J.~G.~Andrews, ``Heterogeneous cellular networks with flexible cell association: A comprehensive downlink SINR analysis,'' \emph{IEEE Trans. Wireless Commun.}, vol.~11, no.~10, pp. 3484--3495, Oct. 2012.
\bibitem{CM2017an}
J.~An, K.~Yang, J.~Wu, N.~Ye, S.~Guo, and Z.~Liao, ``Achieving sustainable ultra-dense heterogeneous networks for 5G,'' \emph{IEEE Commun. Mag.}, vol.~55, no.~12, pp.~84--90, Dec. 2017.
\bibitem{alzer1997some}
H.~Alzer, ``On some inequalities for the incomplete Gamma function,''\emph{Mathematics of computation of the American mathematical society}, vol.~66, no. 218, pp. 771--778, Apr. 1997.
\bibitem{JSAC2014Akdeniz}
M.~R.~Akdeniz, Y.~Liu, M.~K.~Samimi, S.~Sun, S.~Rangan, T.~S. Rappaport, and E.~Erkip, ``Millimeter wave channel modeling and cellular capacity evaluation,'' \emph{IEEE J. Sel. Areas Commun.}, vol.~32, no.~6, pp. 1164--1179, Jun. 2014.

\end{thebibliography}

\end{document}